\documentclass[final]{elsarticle}

\usepackage[utf8]{inputenc} 
\usepackage[T1]{fontenc}    
\usepackage{hyperref}
\usepackage{overpic}
\usepackage{amsmath}
\usepackage{amssymb}
\usepackage{utfsym}
\usepackage{bbm}
\usepackage{threeparttable}
\usepackage{color,xcolor}

\usepackage{geometry}
\journal{Computers in Biology and Medicine}

\begin{document}

\begin{frontmatter}

\title{3D Matting: A Benchmark Study on Soft Segmentation Method for Pulmonary Nodules Applied in Computed Tomography}


\author[1st_address,2nd_address,3th_address]{Lin Wang}
\author[1st_address]{Xiufen Ye\corref{mycorrespondingauthor}}
\author[2nd_address]{Donghao Zhang}
\author[3th_address]{Wanji He}
\author[2nd_address,3th_address]{Lie Ju}
\author[4th_address]{Yi Luo}
\author[5th_address]{Huan Luo}
\author[3th_address]{Xin Wang}
\author[2nd_address,3th_address]{Wei Feng}
\author[3th_address]{Kaimin Song}
\author[3th_address]{Xin Zhao}
\author[2nd_address,3th_address]{Zongyuan Ge\corref{mycorrespondingauthor}}
\cortext[mycorrespondingauthor]{Corresponding author\\ 
\textit{
\indent\; Email address: zongyuan.ge@monash.edu (Zongyuan Ge) \\ 
\indent\qquad\qquad\qquad\quad\ yexiufen@hrbeu.edu.cn (Xiufen Ye)\\ 
\indent\qquad\qquad\qquad\quad\ wanglin.mailbox@gmail.com (Lin Wang)}}

\address[1st_address]{College of Intelligent Systems Science and Engineering, Harbin Engineering University, Harbin, China}
\address[2nd_address]{Monash Medical AI Group, Monash University, Clayton, Australia}
\address[3th_address]{Beijing Airdoc Technology Co., Ltd., Beijing, China}
\address[4th_address]{Chongqing Hospital of Traditional Chinese Medicine, Chongqing, China}
\address[5th_address]{Chongqing Renji Hospital of Chinese Academy
of Sciences, Chongqing, China}

\begin{abstract}
Usually, lesions are not isolated but are associated with the surrounding tissues. For example, the growth of a tumour can depend on or infiltrate into the surrounding tissues. Due to the pathological nature of the lesions, it is challenging to distinguish their boundaries in medical imaging. However, these uncertain regions may contain diagnostic information. Therefore, the simple binarization of lesions by traditional binary segmentation can result in the loss of diagnostic information. 
In this work, we introduce the image matting into the 3D scenes and use the alpha matte, i.e., a soft mask, to describe lesions in a 3D medical image. The traditional soft mask acted as a training trick to compensate for the easily mislabelled or under-labelled ambiguous regions. In contrast, 3D matting uses soft segmentation to characterize the uncertain regions more finely, which means that it retains more structural information for subsequent diagnosis and treatment.
The current study of image matting methods in 3D is limited. To address this issue, we conduct a comprehensive study of 3D matting, including both traditional and deep-learning-based methods.
We adapt four state-of-the-art 2D image matting algorithms to 3D scenes and further customize the methods for CT images to calibrate the alpha matte with the radiodensity.
Moreover, we propose the first end-to-end deep 3D matting network and implement a solid 3D medical image matting benchmark. Its efficient counterparts are also proposed to achieve a good performance-computation balance.
Furthermore, there is no high-quality annotated dataset related to 3D matting, slowing down the development of data-driven deep-learning-based methods. To address this issue, we construct the first 3D medical matting dataset. The validity of the dataset was verified through clinicians' assessments and downstream experiments.
The dataset and codes will be released to encourage further research\footnote{Url for codes and dataset: \url{https://github.com/wangsssky/3DMatting}.}.
\end{abstract}


\end{frontmatter}

\section*{Abbreviations and Symbols}{
The following abbreviations are used in this manuscript:\\

\noindent 
{
	\scriptsize
	\begin{tabular}{@{}l|l}
		\hline
		Term	& Description\\
		\hline
		3D 		& Three-dimensional\\
		3DMM 	& 3D Medical Matting network\\
		AUROC 	& Area Under the Receiver Operating Characteristic\\
		CF 		& Closed-Form matting\\
		Conn.	& Connectivity Error\\
		CT   	& Computed Tomography\\
		DL		& Deep-Learning\\
		FLOPs  	& FLoating-point OPerations\\
		Grad.	& Gradient Error\\
		GT 		& Ground Truth\\
		HU 		& Hounsfield Unit\\
		IF 		& Information-Flow matting\\
		KNN 	& KNN matting\\
		LB 		& Learning-Based matting\\
		MRI 	& Magnetic Resonance Imaging\\
		MSE		& Mean Squared Error\\
		PET 	& Positron Emission Tomography\\
		SAD		& Sum of Absolute Differences\\
		\hline
	\end{tabular}
}

The following symbols are used in this manuscript:\\

\noindent 
{
	\scriptsize
	\begin{tabular}{@{}l|l}
		\hline
		Term	& Description\\
		\hline
		$\boldsymbol{\alpha}$	& alpha matte\\
		$\mathcal{I}$			& 2D image\\
		$\boldsymbol{I}$		& identity matrix\\
		$\mathcal{B}$			& background\\
		$\mathcal{D}$			& diagonal matrix indexing the position of the constraints\\
		$\mathcal{F}$			& foreground\\
		$J(\cdot)$				& cost function\\
		$\mathcal{L}$			& loss\\
		$\boldsymbol{L}$		& matting Laplacian matrix\\
		$\mathcal{M}$			& binary mask\\
		$\mathcal{R}_b$			& background region\\ 
		$\mathcal{R}_f$			& foreground region\\		
		$\mathcal{R}_u$			& unknown region\\
		$\mathcal{S}$ 			& constraints of traditional matting methods\\
		$\mathcal{V}$			& 3D volume\\
		$\delta(\cdot)$			& kronecker delta\\
		$\epsilon$				& a coefficient of regularizer\\
		$\lambda$				& weight of the constraints\\
		$\eta,\theta$ 			& loss weighting coefficients\\
		$\mu$					& mean\\
		$\sigma$				& variance\\
		$w$ 					& local window\\

		\hline
	\end{tabular}
}
}

\section{Introduction}
\begin{figure}[!t]
	\begin{center}
		\begin{overpic}[width=\textwidth]{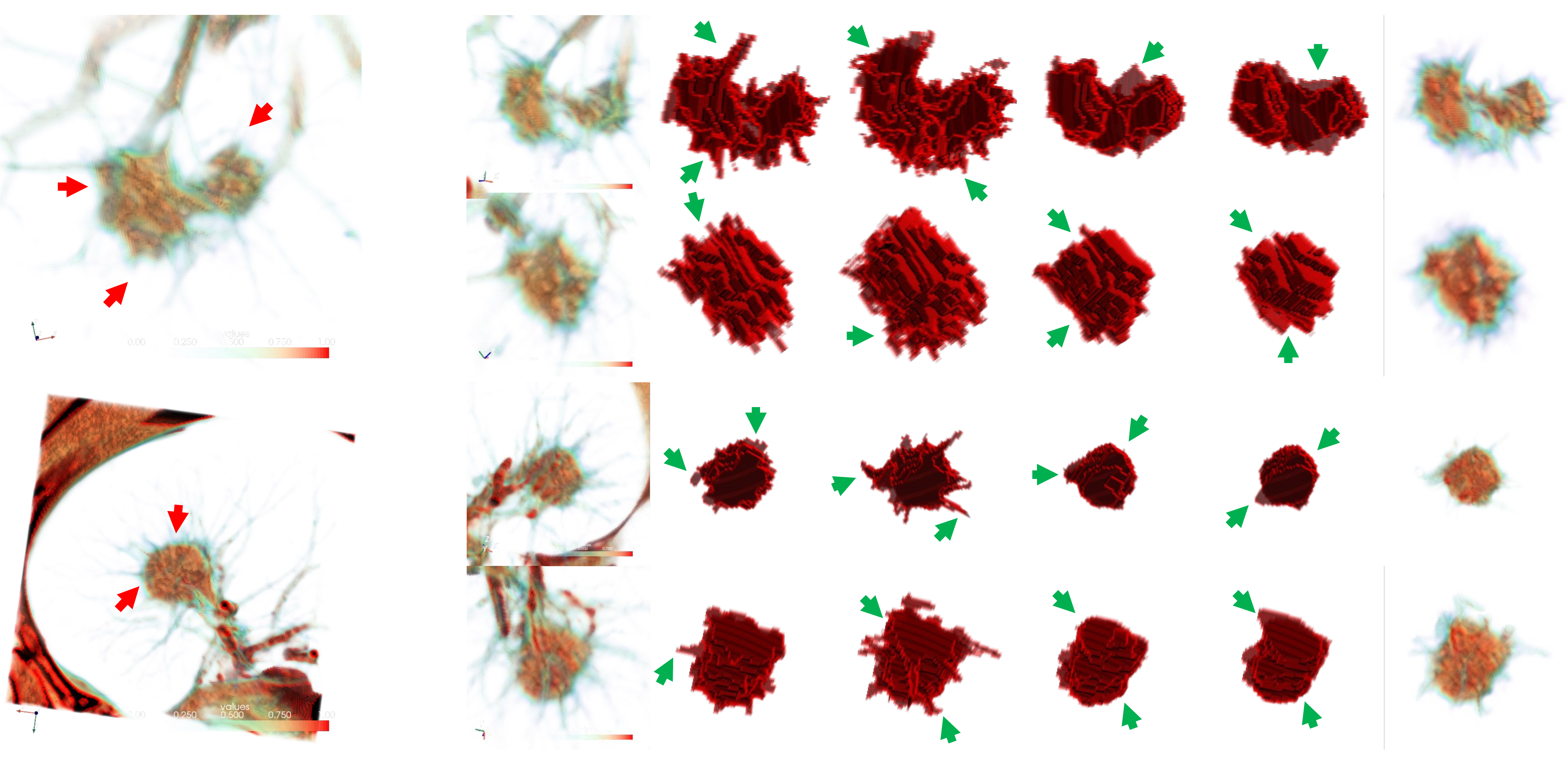}
			\put(34,-3.5){(a)}
			\put(45.5,-3.5){(b)}
			\put(57.5,-3.5){(c)}
			\put(69,-3.5){(d)}
			\put(80.5,-3.5){(e)}
			\put(92,-3.5){(f)}
		\end{overpic}
	\end{center}
	\caption{
		Examples of describing lung nodules~\cite{armato2004lung} with binary masks and alpha mattes. 
		We show two pulmonary nodule samples (Left) in multiple views (Right).
		Due to the blurred boundaries of the nodules (indicated by red arrows), it is challenging for the binary masks labelled by different clinicians (b)-(e) to achieve agreement (indicated by green arrows).
		Moreover, manual labelling of lesions is done slice by slice, making it difficult to maintain structural continuity of lesions between slices.
		However, the ground truth soft masks (f), i.e., the alpha mattes, have a better ability to represent the details of the lesions, with better continuity in between the slices.
		The redder the color, the more likely it is to be part of a lesion.}
	\label{fig:fuzziness}
\end{figure}
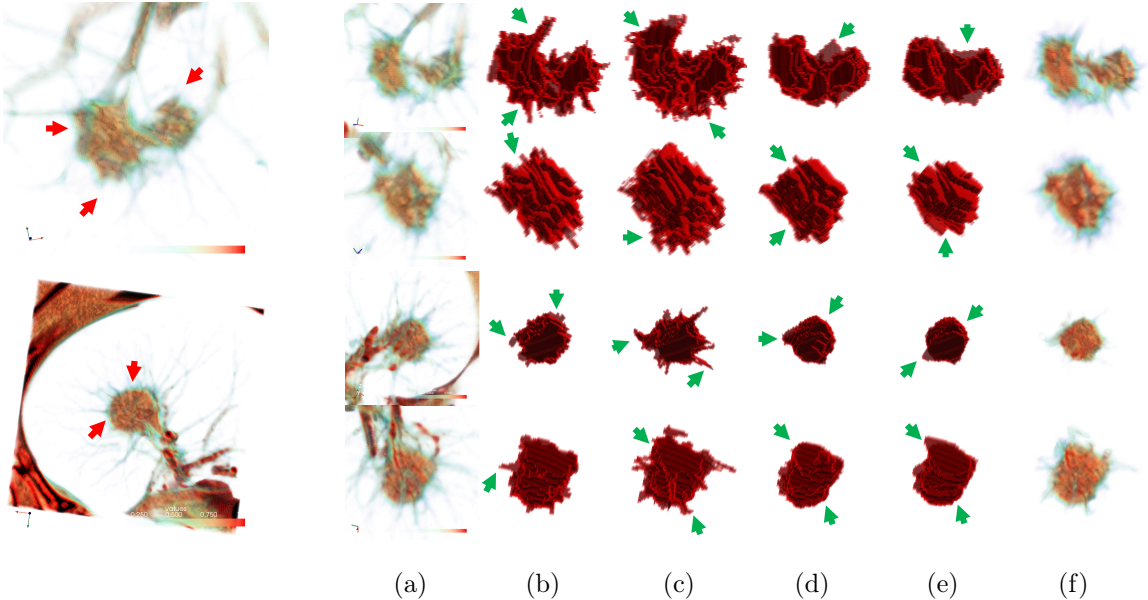

Due to image noises, the occlusion of human tissues, the principles of medical imaging, and the anatomical structure characteristics of lesions, fuzzy boundaries are almost inevitable and ubiquitous in medical images~\cite{kohl2019hierarchical,wang2021medical,li2021applications,elbatel2022mammograms}.
Binary masks are the most commonly used to describe diseased areas in segmentation tasks. However, the fuzzy boundary hinders the accurate recognition of the lesion area, which results in uncertainty in the labelling process, and adversely affects diagnoses and treatments downstream. Figure~\ref{fig:fuzziness} shows some examples of fuzziness in medical images. The morphology of the uncertain regions around the pulmonary nodules is hard to be described by the binarized boundaries with binary masks, which may result in the loss of information.

Many studies have been put forward to reduce the influence of ambiguity in medical image segmentation. Multi-annotated datasets are proposed to reduce labelling bias~\cite{armato2004lung,codella2018skin}, and probabilistic models attempt to describe the lesion distribution~\cite{kohl2018probabilistic,baumgartner2019phiseg,kohl2019hierarchical,gantenbein2020revphiseg}, etc. 
However, due to the complicated reasons for ambiguity, the elimination of ambiguity is still challenging.

Instead of identifying perfect lesion boundaries that are difficult to annotate using binary-style masks, medical matting makes use of the information contained in the ambiguous regions~\cite{wang2021medical}. 
It introduces the image matting technique into the field of medicine, and regards medical images as a mixture of lesions and healthy tissues. 
This mixing factor is known as \textit{alpha matte}.
In such a way, the fuzzy boundary area is regarded as the transition region from pure lesions to pure healthy tissues. 
The alpha matte can be used as a soft mask to describe the anatomical structures of lesions more comprehensively than a binary mask.

In addition, the fuzzy areas contain important diagnostic information.
For example, in thoracic CT images, the fuzzy area around lung nodules in the lung CT images may refer to two kinds of borders, the indistinct border and amorphous ground-glass shadow, which are vital for the clinical staging of nodules~\cite{dyer2020implications}.
Therefore, by keeping more details, medical matting provides more information for downstream tasks (such as nodule grading and precise radiotherapy) than binary segmentation.

The previous works focused on 2D medical images. The 3D medical image matting research is very limited in quantity and methodology~\cite{zeng2012region,cheng2017awm,zhao2020improving,fan2018hierarchical,kim2021uacanet,khan2022deep}. 
To the best of our knowledge, only~\cite{zhong20173d,zhong2018improving,liu2020three} have touched upon the problem of the 3D matting, and these methods are all derived from 2D CF~\cite{levin2007closed}. At present, there is no DL-based method that has been investigated.
To address this, we adapt the matting methods to 3D medical image scenes, including four traditional methods and a DL-based method, as more accurate approaches for lesion segmentation and description, especially for fuzzy areas.

As we know, this is the first work to explore the possibility of matting to solve the problem of fuzzy boundaries in 3D medical image segmentation. Furthermore, this is the first attempt to deploy DL-based matting, rather than traditional matting methods, to 3D image data to realize automatic inference without manual intervention.

Firstly, due to the lack of available datasets in the 3D matting scenario, we created a publicly accessible and clinically validated dataset of pulmonary nodules based on the LIDC-IDRI~\cite{armato2004lung}. We hope that it can benefit the 3D matting research community.
Using this dataset, we verify that the alpha matte contains more diagnostic information than the binary mask quantitatively.
Furthermore, four state-of-the-art traditional 2D matting methods are adapted to 3D scenarios and they are further customized to CT, making the dataset built in a semi-automatic approach.
Finally, the 3D medical matting network, a benchmark DL-based 3D model, is proposed as an end-to-end automatic matting network for pulmonary nodules. At the same time, we optimize the benchmark model by simplifying the network structure and introducing the ghost module to achieve a trade-off between performance and computation~\cite{han2020ghostnet}.

Our contributions are summarized as follows:
\begin{itemize}
	\item This is the first comprehensive study of matting methods in 3D medical scenes, especially for lung nodules in CT. Through qualitative clinical evaluation and quantitative downstream experiments, we have verified that alpha matte retains more structural and diagnostic information of lesions than a binary mask.
 	\item The first DL-based trimap-free matting benchmark network (3DMM) is proposed. The 3DMM contains two auxiliary mask branches that predict the overlap and union of the multi-labelled binary masks, providing guidance information for alpha matte prediction. Its inference does not require human participation and is more convenient for real-world applications.
	\item Moreover, we propose several DL-based methods with higher computational efficiency to achieve a better balance between performance and computation for broader applications.
	\item Four state-of-the-art traditional 2D matting methods are adapted to 3D scenes with further improvements to CT images, which provide the methodological basis for the efficient construction of 3D medical matting datasets in a semi-automatic way.
	\item The first clinically validated 3D matting dataset specifically for 3D medical images is proposed and publicly available to the research community to address the lack of dataset in related studies.
\end{itemize}

The rest of the manuscript is organized as follows:
Section~\ref{sec:related work} provides comprehensive background information on image matting and its applications in medical scenarios. 
Section~\ref{sec:datasets} introduces the 3D matting benchmark dataset. 
Section~\ref{traditional_methods} introduces the traditional 3D matting methods and the optimization specific for CT images.
The trimap-free DL-based methods for 3D medical images are proposed in Section~\ref{methods}.
Section~\ref{experiments} presents the experiments to illustrate the advantages of alpha mattes to binary masks and investigate the performance of DL-based 3D matting methods.
Potential uses and limitations are discussed in Section~\ref{section:discussion}.
Section~\ref{conclusion} is a summary of the whole paper.

\section{Related Work}~\label{sec:related work}
\subsection{Soft Segmentation}
There have been some studies on soft segmentation in the medical field. For example, Kats et al. claimed that pixels around lesions also have diagnostic information and assigned them a soft label in training to improve segmentation performance~\cite{kats2019soft}. Dai et al. used soft masks in data augmentation, which mixed the lesion with the image by using a soft coefficient at the boundaries of the lesions~\cite{dai2022soft}. However, these soft masks can not reflect the structural information of the lesions.

\subsection{Image Matting}
Image matting uses the mixing coefficient $\boldsymbol{\alpha}$, also known as the \textit{alpha matte}, to decompose the image $\mathcal{I}$ to foreground $\mathcal{F}$ and background $\mathcal{B}$, or lesions and its surrounding tissues in medical images~\cite{wang2021medical,aksoy2017designing,cai2019disentangled,chen2013knn,chuang2001bayesian,forte2020fbamatting,levin2007closed,lutz2018alphagan,wang2008image,xu2017deep}. It can be defined as:
\begin{equation}
	\mathcal{I}_i = \boldsymbol{\alpha}_i \mathcal{F}_i + (1-\boldsymbol{\alpha}_i) \mathcal{B}_i.
	\label{eq:image_matting}
\end{equation}
Image matting is a particular type of image segmentation that uses the alpha matte, a soft mask, to describe the target. It is beneficial for image/video editing when dealing with the blurred boundaries. 
Compared with binary masks, alpha mattes can better depict lesions with more details~\citep{wang2021medical}.

There are four terms in Eq.~\ref{eq:image_matting}. However, only $\mathcal{I}$ is known.
Therefore, the matting is an ill-posed problem and is challenging to be solved directly~\cite{yao2017comprehensive}.
A common practice is to reduce the problem complexity by introducing a prior map called \textit{trimap} as constraints, indexing the regions of the foreground $\mathcal{R}_f$, background $\mathcal{R}_b$, and unknown $\mathcal{R}_u$.
Therefore, according to whether the trimap is used, the matting methods can be categorised into trimap-based and trimap-free methods. Compared with trimap-free methods, the trimap-based methods generally achieve better performance as the trimap provides additional information. However, generating the trimap requires extra-manual labour and limits practical application~\cite{chen2022pp}.

According to the usage of DL techniques, image matting methods can be divided into traditional methods~\cite{aksoy2017designing,chen2013knn,chuang2001bayesian,levin2007closed} and DL-based methods~\cite{cai2019disentangled,forte2020fbamatting,lutz2018alphagan,xu2017deep}. 
To the best of our knowledge, almost all of the traditional matting methods are trimap-based, which requires a trimap or scribbles that provide prior information of the foreground and background. Such information is used to infer the values of alpha matte in the uncertain areas by deriving pre-defined rules or assumptions. For instance, the CF assumes that the foreground and background are locally smooth and that the image in a small window can be represented by them with a linear function~\cite{levin2007closed}.

Due to the latest developments in artificial intelligence, the DL-based methods, driven by extensive data, tend to have better performance and a more comprehensive range of applications than traditional methods. 
Moreover, the trimap-free methods become possible as the prior information can be learned by the context-aware networks, which makes matting automatic and easy to use~\cite{2016Deep, chen2022pp}.
However, the traditional methods are still popular because they require no tedious training and thus have better generalization in new applications with insufficient data. 
The readers can refer to~\cite{wang2008image} for a more comprehensive understanding of image matting.

\subsection{Matting in Medical Applications}

Image matting can produce fine boundaries as it infers the uncertain regions by the prior information from foregrounds and backgrounds. 
Therefore, most image matting applications in medical scenarios regarded it as a post-processing method to improve the segmentation performance.
Zeng et al. refined edges of the segmentation results by using the CF~\citep{levin2007closed} directly on each slice of the PET volume data~\citep{zeng2012region}. The boundary pixels were regarded as a mixture of the foreground (tumours) and background (normal tissue).
Cheng et al. deployed an optimized CF~\citep{levin2007closed} on medical images of different imaging modalities, which improved the segmentation performance~\citep{cheng2017awm}.
Zhao et al. utilized the segmentation result to create a trimap via the bi-level thresholding and improved the segmentation performance of the uncertain regions further by minimizing a local matting loss~\citep{zhao2020improving}.

Different from the methods mentioned above, Medical Matting introduced the alpha matte as an alternative to the binary mask, which was more expressive and could describe the lesion in more detail. Moreover, the alpha matte reflected the uncertainty of the lesion property, expanding the scope of matting in medical scenes~\cite{wang2021medical}.

\subsection{Matting in 3D Scenarios}
Matting in 3D is not well studied as the majority of the natural images are 2D images. However, a large proportion of the images of various modalities in medical scenarios are in 3D, therefore there is a strong demand for 3D matting-based methods to be proposed in medical applications. 
The research on 3D medical matting is limited. All of them are based on traditional methods and are used as an auxiliary to binary segmentation. 
For example, Zhong et al. adapted CF~\cite{levin2007closed} to 3D, and used alpha mattes as probability maps of tumours in calculating the region cost for PET-CT co-segmentation~\cite{zhong20173d,zhong2018improving}.
Liu et al. also used a 3D CF for organ model extraction for Virtual Human Project with significant efficiency improvement~\cite{liu2020three}.

Most existing methods still aim at obtaining a better binary mask but do not touch the root cause of the matting problem, that is, using soft masks to describe the lesions in more detail. In this manuscript, by contrast, the matting is used to generate a depictive representation of the lesions.
Compared with the 2D matting methods applied in 3D scenes, we provide a series of native 3D matting methods rather than a combination of slice-by-slice processing by 2D matting methods. Therefore, the structural continuities among different slices can be maintained.
Compared with the important role of 3D data in medical diagnosis, the current research on 3D matting is not comprehensive and in-depth.
In this work, we investigate the traditional matting methods in 3D medical images and provide a trimap-free solution based on DL.

\section{Benchmark Dataset}~\label{sec:datasets}
Although the DL-based methods are the frontier of 2D matting research nowadays and usually achieve better performance, they require a relatively large number of samples for training~\cite{cai2019disentangled,forte2020fbamatting,lutz2018alphagan,xu2017deep}. 
However, because the matting of 3D scenes has not been well studied, there are no available datasets in the research community of 3D medical matting . 
The lack of datasets severely limits the research of 3D medical matting.
In order to conduct quantitative research on 3D matting and contribute to the research community, we established a 3D matting dataset of CT images of lung nodules based on the LIDC-IDRI dataset~\cite{armato2004lung}.

\subsection{Challenges of Manual Labelling}
Manual labelling is time-consuming and labour-intensive, which is especially true for 3D data. 
Generally speaking, an experienced radiologist needs at least 15 minutes to label a 2D image, which is almost unaffordable for labelling a 3D medical image with thousands of slices.
Moreover, the slice-by-slice labelling may lead to discontinuity between neighbouring slices.
Therefore, we adopt the method of alpha matte labelling in 2D images~\cite{2016Deep}, combined with manual screening to achieve semi-automation of labelling and improve labelling efficiency while maintaining labelling quality.

\subsection{Labelling Pipeline and Quality Control}
For 2D portrait matting, Shen~et al. deployed the CF~\cite{levin2007closed} and KNN~\cite{chen2013knn} to generate alpha mattes and then selected the better one describing the portrait as the GT~\cite{2016Deep}. 
To set up the research for our 3D matting task, four state-of-the-art traditional 2D matting methods, which are adapted into 3D scenes and optimized for CT images (introduced in Section~\ref{traditional_methods}), are employed to generate a set of alpha mattes for each 3D nodule image. 
Then we manually choose the one which depicts the nodule the best as the GT label (See Figure~\ref{fig:fuzziness}~(f) as an example.). 
However, the corresponding nodule image will be discarded if all the generated alpha mattes cannot meet our quality requirement. 

\begin{figure}[!t]
	\begin{center}
		\includegraphics[width=\textwidth]{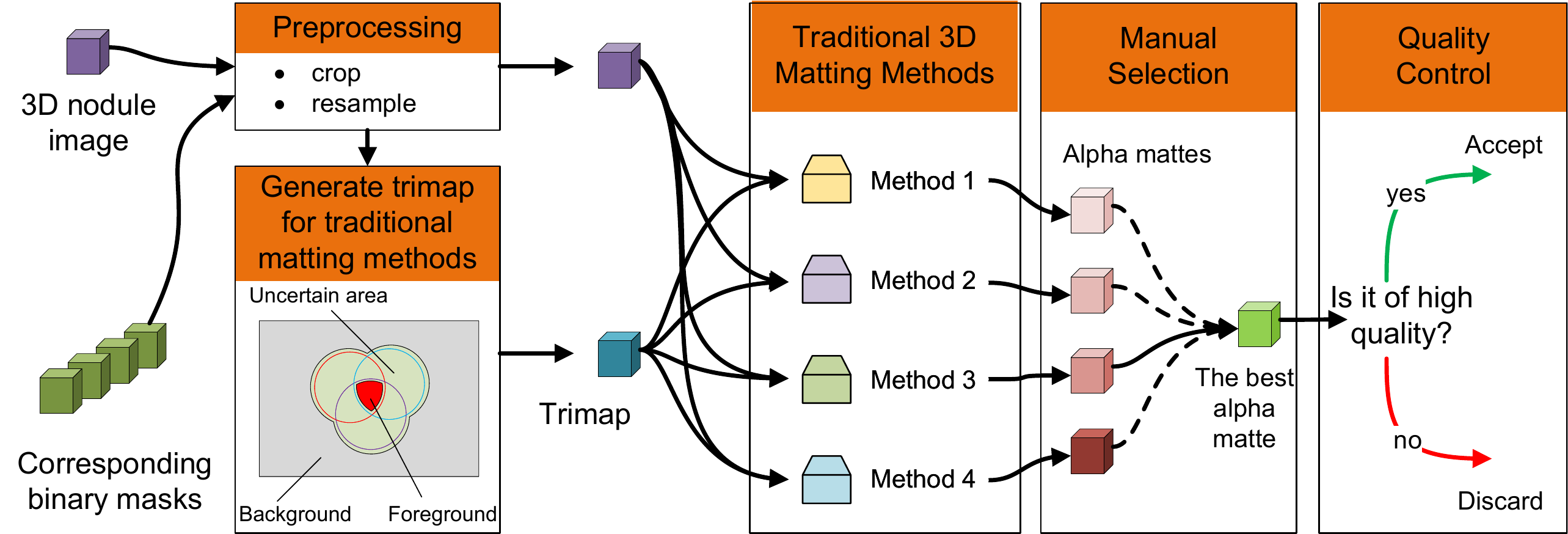}
	\end{center}
	\caption{
		The schematic diagram of labelling pipeline.
		The dataset is generated in a semi-automatic way, using the customized traditional 3D matting methods and multiple binary annotations. The quality control keeps the dataset to a high-quality standard.
	}
	\label{fig:labelling_pipeline}
\end{figure}

Specifically, for preprocessing, each nodule is centred and cropped transversely with the size of $128\times 128$ and sagittally with 3-slice padding at both ends. 
Then the spacing of these 3D patches is resampled to be $0.5 mm$ in each dimension. 
The traditional methods for creating the alternative alpha mattes need no training but only provide the definite foreground, background, and uncertain areas as constraints (i.e., trimap). In the 2D scenario, such auxiliary maps are annotated manually. In this work, we utilize the information of the multi-annotated binary masks to reduce the workload further.
Since each nodule is labelled by four binary masks, we take the intersection of the four binary masks as the foreground, the complement of the union as the background, and the remaining as the unknown region.
Practically, we dilate the unknown region to reduce the bias of manual labelling.
Finally, after manual quality control filtering, we obtain the 3D medical matting dataset, which includes 864 3D nodule images from 542 patients.
Figure~\ref{fig:labelling_pipeline} illustrates the pipeline of the annotation.
Experiments in Section~\ref{Exp_Trad_3D_Matting} illustrate the validity of the dataset quantitatively.

\subsection{Clinicians' Evaluation}
\begin{figure}[!t]
	\begin{center}
		\includegraphics[width=0.4\textwidth]{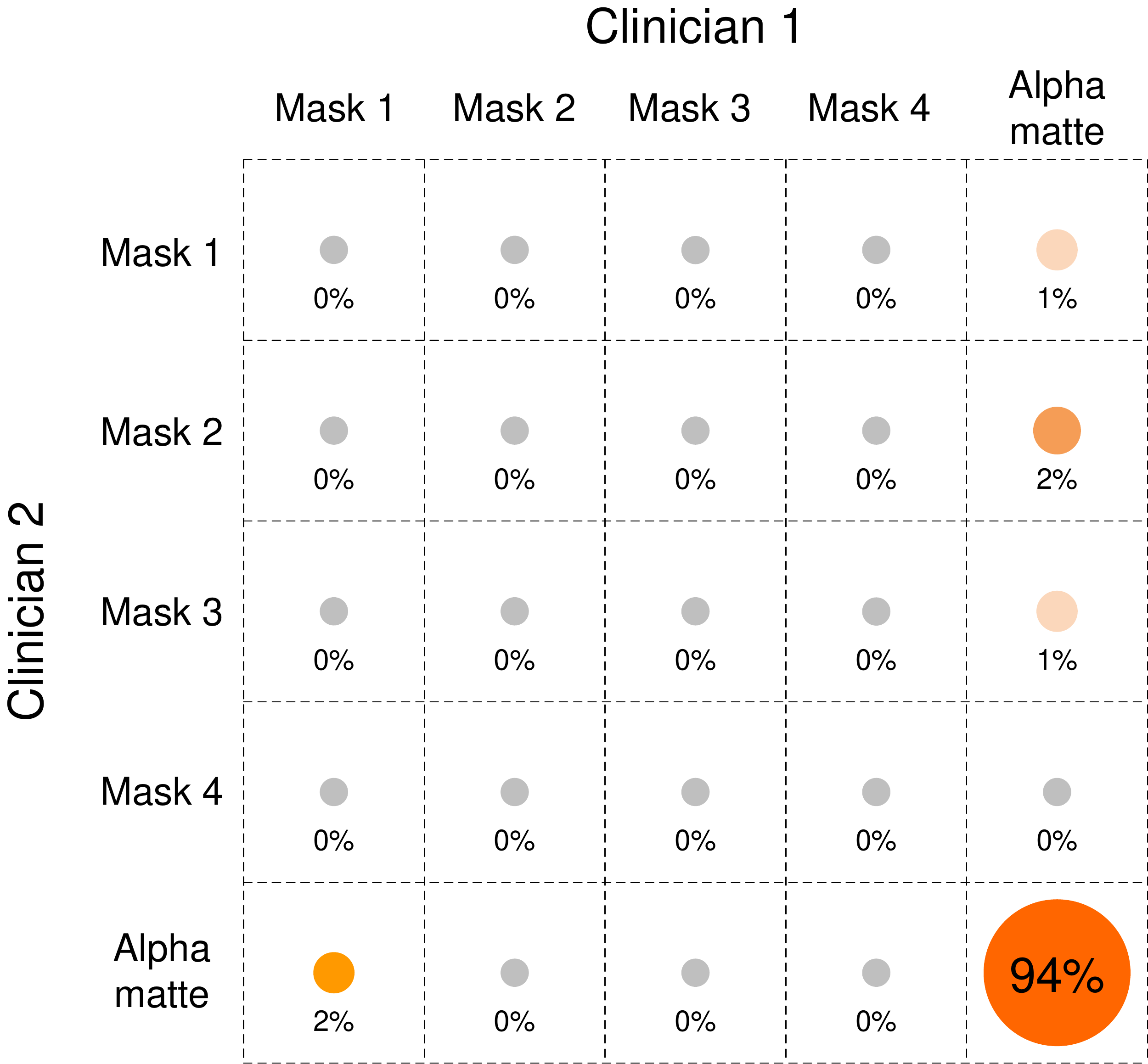}
	\end{center}
	\caption{
		The confusion matrix of the evaluation results shows the clinicians' preference for the alpha matte labels rather than the binary masks.
	}
	\label{fig:clinicians_matrix}
\end{figure}

To further verify the proposed dataset, two clinicians are invited to judge the GT alpha mattes. 
One hundred samples are randomly selected from the customized dataset. 
Clinicians need to select the label that most accurately represents the 3D lesion in each sampled image from the four manual binary masks and our GT alpha matte, in accordance with their clinical experience.
The evaluation results are shown in Figure~\ref{fig:clinicians_matrix}.
Using an alpha matte as a lesion annotation is favoured by clinicians, because it can describe lesions more accurately and finely.


\section{Traditional Matting Method Adaptation to 3D Scenarios}\label{traditional_methods}

\subsection{3D Traditional Matting Methods}

According to the online matting benchmark\footnote{Url for the online matting benchmark: \color{magenta}{\url{http://alphamatting.com}}.}, four state-of-the-art matting methods, i.e., the CF~\cite{levin2007closed}, KNN~\cite{chen2013knn}, LB~\cite{zheng2009learning}, and IF~\cite{aksoy2017designing}, have been selected as the basis of our research because they rank first among the traditional matting methods.
According to the taxonomy in~\cite{yao2017comprehensive}, CF and LB are affinity-based methods that construct the relationship between adjacent pixels (e.g., pixels defined in a local window) first and then propagate the label information from $\{\mathcal{R}_f,\mathcal{R}_b\}$ to $\mathcal{R}_u$. KNN and IF are hybrid methods that use both information from local pixels and non-local pixels. 

Following these traditional 2D solutions, we expand the space from 2D into 3D. 
In general, the solution of alpha matte in these methods can be written as the function:

\begin{equation}
	\boldsymbol{\alpha} = f\left(\mathcal{V}, \mathcal{S}, \mathcal{D},\lambda\right).
\end{equation}
$\mathcal{V}$ is the input 3D volume containing $N$ voxels.
$\mathcal{S}$ is a $N\times 1$ matrix, providing the prior restriction information $\mathcal{R}_f$ and $\mathcal{R}_b$. $\mathcal{D}$ is a $N\times N$ diagonal matrix indexing the position of the restriction, and $\lambda$ is the weight of the restriction information.
In the case of 3D CF, the alpha matte is computed as:
\begin{equation}
	\begin{aligned}
		\boldsymbol{\alpha}&=\mathit{arg}\, \underset{\boldsymbol{\alpha}}{min}\ 
		\boldsymbol{\alpha}^T \boldsymbol{L} \boldsymbol{\alpha}+\lambda(\boldsymbol{\alpha}-\mathcal{S})^T\mathcal{D}(\boldsymbol{\alpha}-\mathcal{S}),\\
		\boldsymbol{L}(p,q)&=\sum_{n|(i,j) \in w_n}
		\left(
			\delta_{i,j}-\frac{1}{k^3}(\frac{1}{\sigma_n^2 + \frac{\epsilon}{k^3}}(\mathcal{V}_{ni}-\mu_n)(\mathcal{V}_{nj}-\mu_n) + 1)
		\right).
	\end{aligned}
\end{equation}
$\boldsymbol{L}$ is a $N\times N$ coefficient matrix, i.e., \textit{matting laplacian matrix}.
$\boldsymbol{L}(p,q)$ is the $(p,q)$-th entry of $\boldsymbol{L}$, and $(i,j)$ is the corresponding entry of $(p,q)$ in $k \times k \times k$ size local window of voxels $w_n$.  
$\mathcal{V}_{n(\cdot)}$ is the $(\cdot)$-th voxel in $w_n$ of $\mathcal{V}$. 
$\delta_{i,j}$ is the Kronecker delta~\cite{weisstein2002kronecker}. $\mu_n$ and $\sigma_n^2$ are the mean and variance of $\mathcal{V}$ in $w_n$, while $\epsilon$ is a coefficient of the regularizer. 
The details of the derivation is provided in Section~\ref{3DCF_Derivation}.

\subsubsection{Derivation of 3D closed-form matting}\label{3DCF_Derivation}
Inspired by 2D CF, we provide a derivation of the 3D CF as follows~\cite{levin2007closed,zhong20173d,liu2020three}.
Formally, the 3D CF takes a volume $\mathcal{V}$, such as CT slices or MRI images, as input, which is assumed to be a composite of a foreground volume $\mathcal{F}$ (e.g., lesions, target organs) and a background volume $\mathcal{B}$ (healthy tissues).

\begin{equation}
\mathcal{V}_i=\boldsymbol{\alpha}_i\mathcal{F}_i+(1-\boldsymbol{\alpha}_i)\mathcal{B}_i,
\end{equation}
where $\boldsymbol{\alpha}_i$ is the voxel's foreground opacity.

As $\mathcal{F}$ and $\mathcal{B}$ are anatomical structures, we can assume the $\mathcal{F}$ and $\mathcal{B}$ are locally smooth. Hence, we have: 
\begin{equation}
    \boldsymbol{\alpha}_i\approx \boldsymbol{a}\mathcal{V}_i+\boldsymbol{b}, \forall i \in w,
\end{equation}
where $\boldsymbol{a}=\frac{1}{\mathcal{F}-\mathcal{B}}, \boldsymbol{b}=-\frac{\mathcal{B}}{\mathcal{F}-\mathcal{B}}$ and $w$ is a small 3D window.

The goal is to find a set of $\boldsymbol{\alpha}$, $\boldsymbol{a}$, $\boldsymbol{b}$ to minimize the cost function:
\begin{equation}
    J(\boldsymbol{\alpha}, \boldsymbol{a}, \boldsymbol{b})=\sum_{j\in \mathcal{V}}\left( \sum_{i\in w_j}\left( \boldsymbol{\alpha}_i-\boldsymbol{a}_j\mathcal{V}_i-\boldsymbol{b}_j\right)^2 + \epsilon \boldsymbol{a}_j^2 \right),
\end{equation}
where $w_j$ is a small window around voxel $j$. $\epsilon$ is a coefficient of regularizer.

Suppose we have $N$ voxels in the volume and the window size is set to $k\times k\times k$. Then the cost function can be written as:
\begin{equation}
\begin{aligned}
J(\boldsymbol{\alpha}, \boldsymbol{a}, \boldsymbol{b})&=\sum_{n=1}^{N} \left( \sum_{i=1}^{k^3}\left( \boldsymbol{\alpha}_{ni} - \boldsymbol{a}_n\mathcal{V}_{ni}-\boldsymbol{b}_n \right)^2 + \epsilon \boldsymbol{a}_n^2\right)\\
&=\sum_{n=1}^{N} \left( \left|
\begin{bmatrix} 
    \boldsymbol{a}_n\mathcal{V}_{n1}+\boldsymbol{b}_n\\ \vdots\\ \boldsymbol{a}_n\mathcal{V}_{nk^3}+\boldsymbol{b}_n \\ \sqrt{\epsilon}\boldsymbol{a}_n 
\end{bmatrix} -
\begin{bmatrix}
    \boldsymbol{\alpha}_{n1}\\ \vdots\\ \boldsymbol{\alpha}_{nk^3} \\0 
\end{bmatrix}
\right|^2\right)\\
&=\sum_{n=1}^{N} \left( \left|
\begin{bmatrix} 
    \mathcal{V}_{n1}& 1\\ \vdots\\ \mathcal{V}_{nk^3}&1 \\ \sqrt{\epsilon}& 0 
\end{bmatrix}
\begin{bmatrix} 
    \boldsymbol{a}_n\\ \boldsymbol{b}_n
\end{bmatrix} -
\begin{bmatrix}
    \boldsymbol{\alpha}_{n1}\\ \vdots\\ \boldsymbol{\alpha}_{nk^3} \\0 
\end{bmatrix}
\right|^2\right).
\end{aligned}
\end{equation}

Set $\boldsymbol{G}_n=\begin{bmatrix} 
    \mathcal{V}_{n1}& 1\\ \vdots &\vdots\\ \mathcal{V}_{nk^3}&1 \\ \sqrt{\epsilon}& 0 
\end{bmatrix}
$, 
$
\overline{\boldsymbol{\alpha}}_n=\begin{bmatrix}
    \boldsymbol{\alpha}_{n1}\\ \vdots\\ \boldsymbol{\alpha}_{nk^3} \\0 
\end{bmatrix}
$, 
$
\boldsymbol{c}_n=\begin{bmatrix} 
    \boldsymbol{a}_n\\ \boldsymbol{b}_n
\end{bmatrix} 
$, we have
\begin{equation}
    J(\boldsymbol{\alpha},\boldsymbol{a},\boldsymbol{b})=\sum_{n=1}^{N} \left|\boldsymbol{G}_n\boldsymbol{c}_n - \overline{\boldsymbol{\alpha}}_n\right|^2.
	\label{Eq:J_alpha_ab}
\end{equation}

We denote $\boldsymbol{c}_n^*$ as the optimal pair $\boldsymbol{a}_n^*$ and $\boldsymbol{b}_n^*$ in each window $w_n$. 
Take the partial derivative of $\left|\boldsymbol{G}_n\boldsymbol{c}_n - \overline{\boldsymbol{\alpha}}_n\right|^2$ with respect to $\boldsymbol{c}_n$ and make it equal to 0.
\begin{equation}
    \frac{\partial \left|\boldsymbol{G}_n\boldsymbol{c}_n - \overline{\boldsymbol{\alpha}}_n\right|^2}{\partial \boldsymbol{c}_n}=
    2\boldsymbol{G}_n^T(\boldsymbol{G}_n\boldsymbol{c}_n-\overline{\boldsymbol{\alpha}}_n)=0
\end{equation}

Then we get $\boldsymbol{c}_n^* = (\boldsymbol{G}_n^T\boldsymbol{G}_n)^{-1}\boldsymbol{G}_n^T\overline{\boldsymbol{\alpha}}_n$ and subtitute it into Equation~\ref{Eq:J_alpha_ab}, and we have

\begin{equation}
\begin{aligned}
J(\boldsymbol{\alpha}) 
&=\sum_{n=1}^{N} \left|(\boldsymbol{I}-\boldsymbol{G}_n(\boldsymbol{G}_n^T\boldsymbol{G}_n)^{-1}\boldsymbol{G}_n^T)\overline{\boldsymbol{\alpha}}_n\right|^2.
\end{aligned}
\label{Eq:Ja}
\end{equation}

Denote $\boldsymbol{I}-\boldsymbol{G}_n(\boldsymbol{G}_n^T\boldsymbol{G}_n)^{-1}\boldsymbol{G}_n^T$ as $\overline{\boldsymbol{G}}_n$, the Equation~\ref{Eq:Ja} can be written as:
\begin{equation}
\begin{aligned}
    J(\boldsymbol{\alpha})&=\sum_{n=1}^{N} \left|\overline{\boldsymbol{G}}_n\overline{\boldsymbol{\alpha}}_n\right|^2
    =\sum_{n=1}^{N} (\overline{\boldsymbol{G}}_n\overline{\boldsymbol{\alpha}}_n)^T(\overline{\boldsymbol{G}}_n\overline{\boldsymbol{\alpha}}_n)
    =\sum_{n=1}^{N} \overline{\boldsymbol{\alpha}}_n^T\overline{\boldsymbol{G}}_n^T\overline{\boldsymbol{G}}_n\overline{\boldsymbol{\alpha}}_n.
\end{aligned}
\end{equation}

Because 
\begin{equation}
\begin{aligned}
    \overline{\boldsymbol{G}}_n^T\overline{\boldsymbol{G}}_n &= (\boldsymbol{I}-\boldsymbol{G}_n(\boldsymbol{G}_n^T\boldsymbol{G}_n)^{-1}\boldsymbol{G}_n^T)^T(\boldsymbol{I}-\boldsymbol{G}_n(\boldsymbol{G}_n^T\boldsymbol{G}_n)^{-1}\boldsymbol{G}_n^T)\\
    &= \boldsymbol{I} + (\boldsymbol{G}_n(\boldsymbol{G}_n^T\boldsymbol{G}_n)^{-1})^T\boldsymbol{G}_n^T -(\boldsymbol{G}_n(\boldsymbol{G}_n^T\boldsymbol{G}_n)^{-1})^T\boldsymbol{G}_n^T -\boldsymbol{G}_n(\boldsymbol{G}_n^T\boldsymbol{G}_n)^{-1}\boldsymbol{G}_n^T\\
    &= \boldsymbol{I} -\boldsymbol{G}_n(\boldsymbol{G}_n^T\boldsymbol{G}_n)^{-1}\boldsymbol{G}_n^T\\
    &= \overline{\boldsymbol{G}}_n,
\end{aligned}
\end{equation}
we have
\begin{equation}
	J(\boldsymbol{\alpha}) =\sum_{n=1}^{N} \overline{\boldsymbol{\alpha}}_n^T\overline{\boldsymbol{G}}_n\overline{\boldsymbol{\alpha}}_n.
\end{equation}

Rewrite the $J(\boldsymbol{\alpha})$ as:
\begin{equation}
J(\boldsymbol{\alpha}) = \boldsymbol{\alpha}^T \boldsymbol{L} \boldsymbol{\alpha},
\end{equation}
where the $\boldsymbol{L}$ is the Laplacian matrix.

We introduce the restriction from the trimap: 
\begin{equation}
\begin{aligned}
    \mathop{min}\limits_{\boldsymbol{\alpha}} \boldsymbol{\alpha}^T\boldsymbol{L}\boldsymbol{\alpha}, 
    \quad s.t. \quad (\boldsymbol{\alpha}-\mathcal{S})^T\mathcal{D}(\boldsymbol{\alpha}-\mathcal{S})=0,
\end{aligned}
\end{equation}
where $\mathcal{S}$ is a $N\times 1$ matrix, providing the restriction information. $\mathcal{D}$ is a $N\times N$ diagonal matrix indexing the position of the restriction.

Therefore, 
\begin{equation}
\begin{aligned}
    J(\boldsymbol{\alpha})&=\boldsymbol{\alpha}^T \boldsymbol{L} \boldsymbol{\alpha}+\lambda(\boldsymbol{\alpha}-\mathcal{S})^T\mathcal{D}(\boldsymbol{\alpha}-\mathcal{S}),
\end{aligned}
\end{equation}
where $\lambda$ is the weight of the restriction information. 

Set $\nabla J(\boldsymbol{\alpha}) = 0$,
\begin{equation}
\begin{aligned}
    \nabla J(\boldsymbol{\alpha}) = 2\boldsymbol{L}\boldsymbol{\alpha} + \lambda(2\mathcal{D}\boldsymbol{\alpha} -2\mathcal{D}\mathcal{S})&=0\\
    (\boldsymbol{L}+\lambda \mathcal{D})2\boldsymbol{\alpha} - 2\lambda \mathcal{D}\mathcal{S} &=0,
\end{aligned}
\end{equation}

\begin{equation}
\begin{aligned}
    \boldsymbol{\alpha} = (\boldsymbol{L}+\lambda \mathcal{D})^{-1}\lambda \mathcal{D}\mathcal{S}.
\end{aligned}
\end{equation}

\begin{equation}
\begin{aligned}
    \boldsymbol{G}_n^T\boldsymbol{G}_n&=
    \begin{bmatrix} 
        \mathcal{V}_{n1}& \cdots& \mathcal{V}_{nk^3}&\sqrt{\epsilon}\\ 1&\cdots&  1& 0 
    \end{bmatrix}
    \begin{bmatrix} 
        \mathcal{V}_{n1}& 1\\ \vdots& \vdots\\ \mathcal{V}_{nk^3}&1 \\ \sqrt{\epsilon}& 0 
    \end{bmatrix}\\
    &=
    \begin{bmatrix} 
        \sum_{i=0}^{k^3} \mathcal{V}_{ni}^2 + \epsilon& \sum_{i=0}^{k^3}\mathcal{V}_{ni}\\ 
        \sum_{i=0}^{k^3} \mathcal{V}_{ni}& k^3 
    \end{bmatrix}
\end{aligned}
\end{equation}

Denote 
\begin{equation}
    \left\{
        \begin{aligned}
            &\mu_n = \frac{1}{k^3}\sum_{i=0}^{k^3}\mathcal{V}_{ni}\\
            &\sigma_n^2 = \frac{1}{k^3}\sum_{i=0}^{k^3}(\mathcal{V}_{ni}-\mu_n)^2
        \end{aligned}
    \right.,
\end{equation}
therefore we have
\begin{equation}
    \left\{
        \begin{aligned}
            &\sum_{i=0}^{k^3}\mathcal{V}_{ni} = {k^3}\mu_n\\
            &\sum_{i=0}^{k^3}\mathcal{V}_{ni}^2 = k^3(\sigma_n^2 + \mu_n^2).
        \end{aligned}
    \right.
\end{equation}

\begin{equation}
\begin{aligned}
    \boldsymbol{G}_n^T\boldsymbol{G}_n&=
    \begin{bmatrix} 
        k^3(\sigma_n^2 + \mu_n^2) + \epsilon& {k^3}\mu_n\\ 
        {k^3}\mu_n& k^3 
    \end{bmatrix}
\end{aligned}
\end{equation}

\begin{equation}
\begin{aligned}
    (\boldsymbol{G}_n^T\boldsymbol{G}_n)^{-1}&=
    \frac{1}{k^3(k^3(\sigma_n^2 + \mu_n^2) + \epsilon)-({k^3}\mu_n)^2}
    \begin{bmatrix} 
        k^3 & -{k^3}\mu_n\\ 
        -{k^3}\mu_n& k^3(\sigma_n^2 + \mu_n^2) + \epsilon
    \end{bmatrix}\\
    &=
    \frac{1}{k^3\sigma_n^2 + \epsilon}
    \begin{bmatrix} 
        1 & -\mu_n\\ 
        -\mu_n& \sigma_n^2 + \mu_n^2 + \frac{\epsilon}{k^3}
    \end{bmatrix}\\
\end{aligned}
\end{equation}

Denote $\frac{1}{k^3\sigma_n^2 + \epsilon}$ as $t_1$, $\sigma_n^2 + \mu_n^2 + \frac{\epsilon}{k^3}$ as $t_2$,

\begin{equation}
\begin{aligned}
    &\quad\overline{\boldsymbol{G}}_n\\
    &=\boldsymbol{I}-\boldsymbol{G}_n(\boldsymbol{G}_n^T\boldsymbol{G}_n)^{-1}\boldsymbol{G}_n^T\\
    &=
    \boldsymbol{I}-t_1
    \begin{bmatrix} 
        \mathcal{V}_{n1}& 1\\ \vdots& \vdots\\ \mathcal{V}_{nk^3}&1 \\ \sqrt{\epsilon}& 0 
    \end{bmatrix}
    \begin{bmatrix} 
        1 & -\mu_n\\ 
        -\mu_n& t_2
    \end{bmatrix}   
    \begin{bmatrix} 
        \mathcal{V}_{n1}& \cdots& \mathcal{V}_{nk^3}&\sqrt{\epsilon}\\ 1&\cdots&  1& 0 
    \end{bmatrix}\\ 
    &= 
    \boldsymbol{I}-t_1
    \begin{bmatrix} 
        \mathcal{V}_{n1}\mathcal{V}_{n1}-\mu_n\mathcal{V}_{n1}-\mu_n\mathcal{V}_{n1}+t_2& \cdots& \mathcal{V}_{n1}\mathcal{V}_{nk^3}-\mu_n\mathcal{V}_{nk^3}-\mu_n\mathcal{V}_{n1}+t_2 & \epsilon-\epsilon\mu_n\\ 
        \vdots& \ddots &\vdots &\vdots\\ 
        \mathcal{V}_{nk^3}\mathcal{V}_{n1}-\mu_n\mathcal{V}_{n1}-\mu_n\mathcal{V}_{nk^3}+t_2& \cdots& \mathcal{V}_{nk^3}\mathcal{V}_{nk^3}-\mu_n\mathcal{V}_{nk^3}-\mu_n\mathcal{V}_{nk^3}+t_2 & \epsilon-\epsilon\mu_n\\ 
        \sqrt{\epsilon}\mathcal{V}_{n1} -\sqrt{\epsilon}\mu_n & \cdots & \sqrt{\epsilon}\mathcal{V}_{nk^3} -\sqrt{\epsilon}\mu_n & \epsilon
    \end{bmatrix}.
\end{aligned}
\end{equation}

\begin{equation}
    \begin{aligned}
        \quad\mathop{\overline{\boldsymbol{G}}_n(i,j)}\limits_{i,j \in w_n}&=
        \delta_{i,j}-t_1(\mathcal{V}_{ni}\mathcal{V}_{nj}-\mu_n\mathcal{V}_{ni}-\mu_n\mathcal{V}_{nj}+t_2)\\
        &=\delta_{i,j}-\frac{1}{k^3\sigma_n^2 + \epsilon}((\mathcal{V}_{ni}-\mu_n)(\mathcal{V}_{nj}-\mu_n) + \frac{k^3\sigma_n^2+\epsilon}{k^3})\\
        &=\delta_{i,j}-\frac{1}{k^3}(\frac{1}{\sigma_n^2 + \frac{\epsilon}{k^3}}(\mathcal{V}_{ni}-\mu_n)(\mathcal{V}_{nj}-\mu_n) + 1),
    \end{aligned}
\end{equation}
where $\delta_{i,j}$ is the Kronecker delta, defined as
\begin{equation}
    \delta_{i,j}=
    \left\{
        \begin{aligned}
            0\quad if\quad i \neq j\\
            1\quad if\quad i =j.
        \end{aligned}
    \right.
\end{equation}

The Laplacian matrix $\boldsymbol{L}$ is a $N\times N$ matrix, whose $(p,q)$-th entry is 
\begin{equation}
    \begin{aligned}
       \sum_{n|(i,j) \in w_n}\left(\delta_{i,j}-\frac{1}{k^3}(\frac{1}{\sigma_n^2 + \frac{\epsilon}{k^3}}(\mathcal{V}_{ni}-\mu_n)(\mathcal{V}_{nj}-\mu_n) + 1)\right),
    \end{aligned}
\end{equation}
where $(i,j)$ is the corresponding entry of $(p,q)$ in window $w_n$.

The solutions of the other traditional methods can refer to the codes. 

\subsection{Optimization for Medical Images}
Different from the common natural images, the manually labelled foreground in medical images may not be a pure foreground or pure lesion tissues. For example, tumours often grow close to healthy tissues or even mix with them. In addition, the mixture contains vital structural information. Simple homogenization of the foreground area will lose potential diagnostic information. Hence, it is inappropriate to set the alpha matte of the aforementioned foreground in medical images to be $1$. 
To address the problems in transferring 2D matting methods directly to 3D, we modify the methods toward the medical images, and use CT as an example.

It is assumed that the areas within a specific HU\footnote{The HU is a quantitative scale for describing radiodensity in CT, which is similar to the intensity in the gray image. It generally ranges from -1000 HU to +~2000 HU for human tissues. In practice, a specific window is selected for better focusing on the particular observing target.}~\cite{denotter2021hounsfield} range $(l_{low}, l_{high})$ in the foreground are pure lesions, and the further away from this range, the lower percentage of lesion tissues. 
Then the foreground of the alpha matte can be written as:
\begin{equation}
	\underset{i\in \mathcal{R}_f}{\boldsymbol{\alpha}_i}= 
	\begin{cases}
	p(\mathcal{V}_i), & l_{high} < \mathcal{V}_i\\
	1, &l_{low} \leq \mathcal{V}_i \leq l_{high} ,\\
	q(\mathcal{V}_i), & \mathcal{V}_i < l_{low}
	\end{cases}
\end{equation}
where $p(\cdot)$ and $q(\cdot)$ are monotonic mappings over a value domain of 0 to 1.
In this way, the alpha matte is associated with the HU, making the alpha matte physically meaningful.
For simplicity but without generality, we set the upper limit of the CT observation window as $l_{low}$ and set $q(\cdot)$ as a linearly increasing function defined over the window width.
By applying the calibrated foreground as a constraint in $\mathcal{S}$, we achieve the optimized 3D traditional matting methods, respectively noted as CF+, KNN+, LB+, and IF+.

For clear description, we refer to the 3D traditional matting methods as \textbf{traditional methods} and their optimized counterparts as \textbf{optimized methods} in the following sections.

\section{3D Medical Matting Network}\label{methods}

\begin{figure}[!t]
	\begin{center}
		\includegraphics[width=\textwidth]{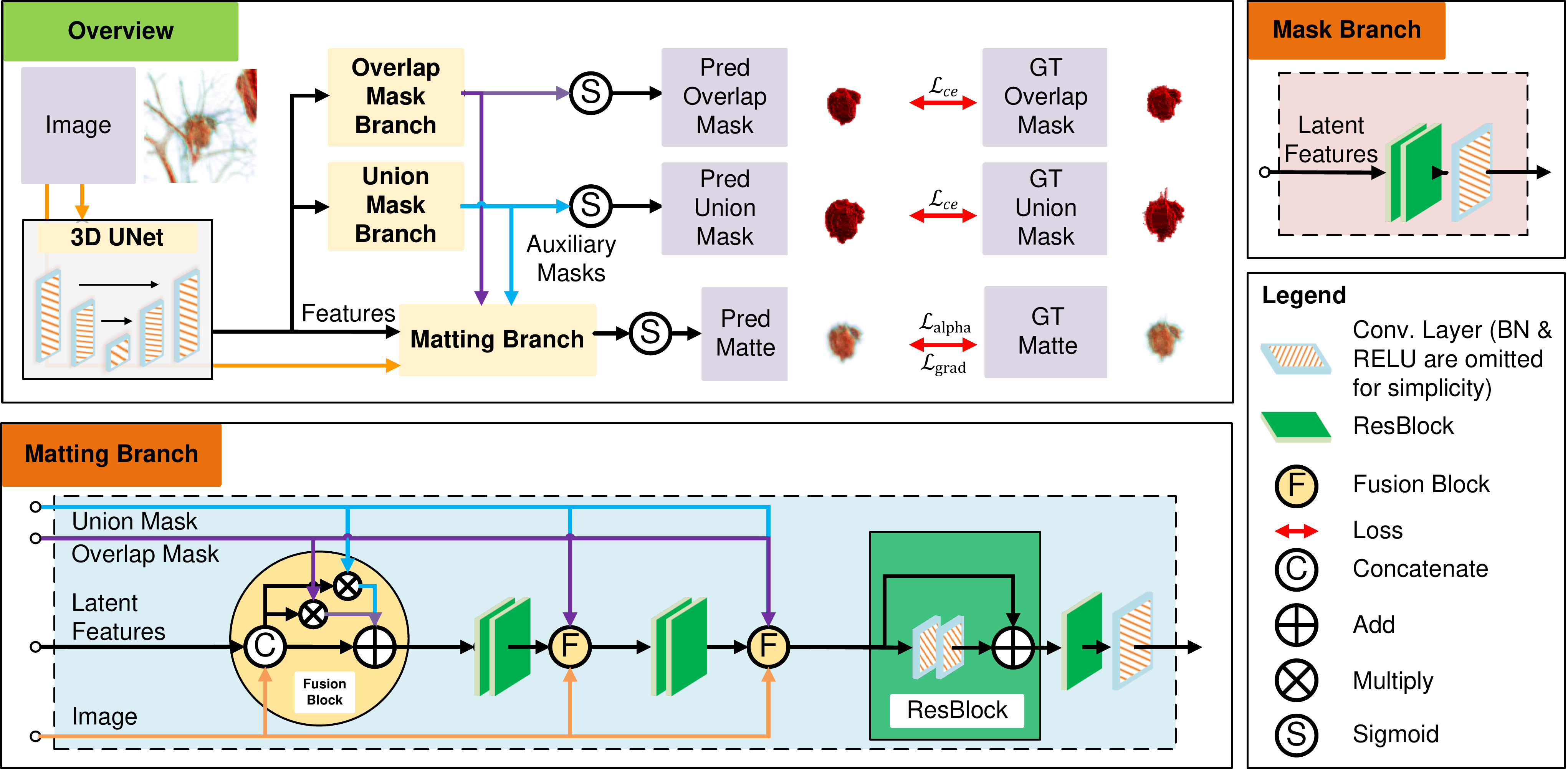}
	\end{center}
	\caption{
		The schematic diagram of the 3D Medical Matting Network.
		As a benchmark model, we keep the network elegant. The network contains three parts, i.e., the 3D UNet, the auxiliary mask branches (overlap/union), and the matting branch. A vanilla 3D UNet is used as the feature extractor. The overlap/union mask branch learns the overlap/union of the manual labelled binary masks as auxiliary masks. The matting branch takes these auxiliary masks as guidance and outputs the predicted alpha matte. In this way, there is no need for the manual input of a priori information, and the process is completely automatic.
	}
	\label{fig:structure}
\end{figure}

The methods proposed in Section~\ref{traditional_methods} do not need model training. 
However, their inferences require human intervention to provide a trimap indexing the foreground and background, which restricts practical applications.
Thus, we propose an end-to-end DL-based model named \textbf{3D} \textbf{M}edical \textbf{M}atting network (\textbf{3DMM}) as a benchmark of 3D matting, which provides a fully automatic matting inference without human involvement.

As the first deep 3D matting network, 3DMM is focused on feasibility verification rather than pursuing high performance.
Therefore, we refer to the successfully applied 2D matting networks, adopt the widely used components and simplify them~\cite{cai2019disentangled,2016Deep,xu2017deep}.
Overall, the 3DMM, a multi-task network\cite{wang2021medical}, consists of four sub-modules, i.e., the \textbf{3D UNet}, \textbf{Overlap / Union mask} branch, and \textbf{Matting} branch.

\subsection{Network Design}
Firstly, the 3D version of vanilla UNet~\cite{ronneberger2015u} is used as the feature extractor for the subsequent three sub-modules.
Secondly, since the matting task usually requires a priori information of the lesion region as a constraint, two auxiliary networks, the Overlap mask branch and the Union mask branch, are proposed to learn the overlap and union of the manual binary masks, providing information of the identified core lesion region and the rough contour, respectively.
Besides being beneficial to matting performance, these two masks also represent the lesion region with different confidence levels, which expand the applicable scenarios.
Their network structures are identical: a group of Resblocks~\cite{he2016deep} followed by a $1\times 1 \times 1$ convolutional layer.
At last, the predicted alpha matte is generated by the Matting branch consisting of three consecutive groups of Resblocks and a $1\times 1\times 1$ convolutional layer.
Before each group of Resblocks, we fuse the auxiliary information obtained from the two mask branches with the latent features.

Overall, the 3DMM is a multi-task network that takes a 3D image as the input and outputs three predictions, i.e., the union mask, the overlap mask, and the alpha matte.
Figure~\ref{fig:structure} provides a schematic view of the 3DMM framework.

\subsection{Losses}
We use the cross-entropy losses $\mathcal{L}_\mathit{ce_{overlap}}$ and $\mathcal{L}_\mathit{ce_{union}}$ for the overlap mask and the union mask prediction, respectively. 
The absolute difference $\mathcal{L}_\mathit{alpha}$ and the gradient difference $\mathcal{L}_\mathit{grad}$ between the predicted alpha matte $\tilde{\boldsymbol{\alpha}}$ and the GT alpha matte $\boldsymbol{\alpha}$ are deployed to the matting branch~\cite{2016Deep,xu2017deep,chen2022pp}.
The definitions are as follows:
\begin{equation}
	\begin{aligned}
		\mathcal{L}_\mathit{mask}&= \mathcal{L}_\mathit{ce_{overlap}} + \mathcal{L}_\mathit{ce_{union}}, \\
		\mathcal{L}_{alpha}&=\frac{1}{\sum\mathbbm{1}_{\tilde{\boldsymbol{\alpha}}}} \sum\nolimits_{i \in \tilde{\boldsymbol{\alpha}}} 
		\left (1 + \mathbbm{1}_{\mathcal{M}_\mathit{union}}(i) \right ) 
		\left \|\tilde{\boldsymbol{\alpha}}_i-\boldsymbol{\alpha}_i\right \|_1, \\
		\mathcal{L}_{\mathit{grad}}&=\frac{1}{\sum\mathbbm{1}_{\tilde{\boldsymbol{\alpha}}}}\sum\nolimits_{i \in \tilde{\boldsymbol{\alpha}}}
		\left (1 + \mathbbm{1}_{\mathcal{M}_\mathit{union}}(i) \right ) 
		\left \|\nabla\tilde{\boldsymbol{\alpha}}_i-\nabla\boldsymbol{\alpha}_i\right \|_1,
	\end{aligned}
	\label{eq:losses}
\end{equation}
where $\mathcal{M}_\mathit{union}$ is the GT union mask used to amplify the weight of the target regions.
Finally, the total loss $\mathcal{L}_\mathit{total}$ can be written as:
\begin{equation}
	\mathcal{L}_\mathit{total}=\mathcal{L}_\mathit{mask} + \eta\mathcal{L}_\mathit{alpha} + \theta\mathcal{L}_\mathit{grad},
	\label{eq:total_loss}
\end{equation}
where $\eta$, $\theta$ are the weight balancing parameters.

\subsection{More efficient models}
\begin{figure}[!t]
	\begin{center}
		\begin{overpic}[width=\textwidth]{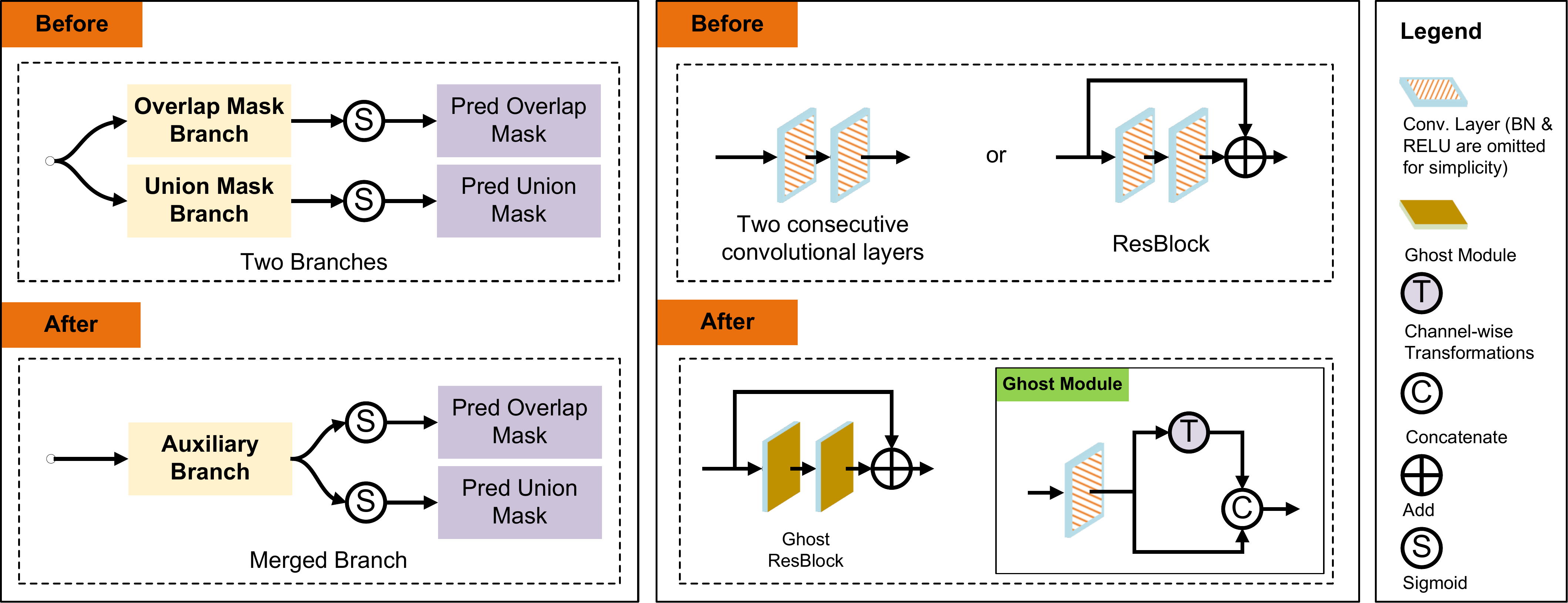}
			\put(18,-5.5){(a)}
			\put(63,-5.5){(b)}	
		\end{overpic}
	\end{center}
	\caption{The schematic diagram shows the modifications made to improve the network efficiency. The optimizations consist of two parts. First, from the point of view of network design, two auxiliary mask branches are merged to reduce the computational complexity, as shown in (a). Then, the ghost modules are used to update the convolutional layers to achieve less memory and computing resources and make the network more efficient (b).}
	\label{fig:efficient_structure}
\end{figure}

Compared to 2D convolutional models, 3D convolutional models tend to have much more parameters and FLOPs, and their practical usability is limited by a lack of computing resources~\cite{kopuklu2019resource}. 
Therefore, in this section, we optimize our 3DMM model to improve the computing efficiency and achieve a good balance between the computing resource and performance.
The modifications are two-fold as shown in Figure~\ref{fig:efficient_structure}: First, the auxiliary mask branches are simplified to reduce the computation from the structural perspective. Then, the efficient ghost module is introduced to make the network more compact~\cite{han2020ghostnet}.

The auxiliary mask branches are established by the overlap and union mask branches. Since they share the same network structure and output similar binary masks, a single auxiliary mask branch with two output channels can be used as an alternative. With this simplification, about half of the computation of the original auxiliary mask branches is saved.

Han et al. observed that redundancy widely exists in intermediate feature maps of the CNN networks~\cite{han2020ghostnet}. On this basis, they built up an efficient and compact module called the ghost module, which consisted of two parts: the traditional convolutional layer, but using fewer filters, and cheap operations to generate more feature maps. 
The traditional convolutional layer produces some base feature maps, while more pseudo feature maps are generated by the channel-wise transformations of the base feature maps, as illustrated in Figure~\ref{fig:efficient_structure}-(b). In practice, the grouped convolution is used to generate cheap feature maps~\cite{krizhevsky2012imagenet}.
In our network, the convolutional ResBlocks and two consecutive convolutional layers unit in each UNet stage are replaced by the Ghost ResBlocks.


\section{Experiments and Results}\label{experiments}

\subsection{Experiment on Traditional 3D Matting Methods}\label{Exp_Trad_3D_Matting}

To verify the reasonability of the GT alpha matte, we compare the ability of various alpha mattes and binary masks to describe lesions by comparing their performance evaluated by the AUROC in the downstream diagnosis of benign/malignant classification of pulmonary nodules. 

The original image, the masks processed in different ways, and the alpha mattes obtained by different methods are used as input, respectively. To illustrate the robustness of the results, we conduct classification experiments on various models of different types and scales, such as Densenet 121, Resnet 34, Resnet 50~\cite{he2016deep,huang2017densely}. 
Each experiment is repeated four times with different seeds. 
Under the same condition, it can be assumed that the more diagnostic information available is in the input, the higher the classification performance will be. 

\begin{figure}[!t]
	\begin{center}
		\begin{overpic}[width=\textwidth]{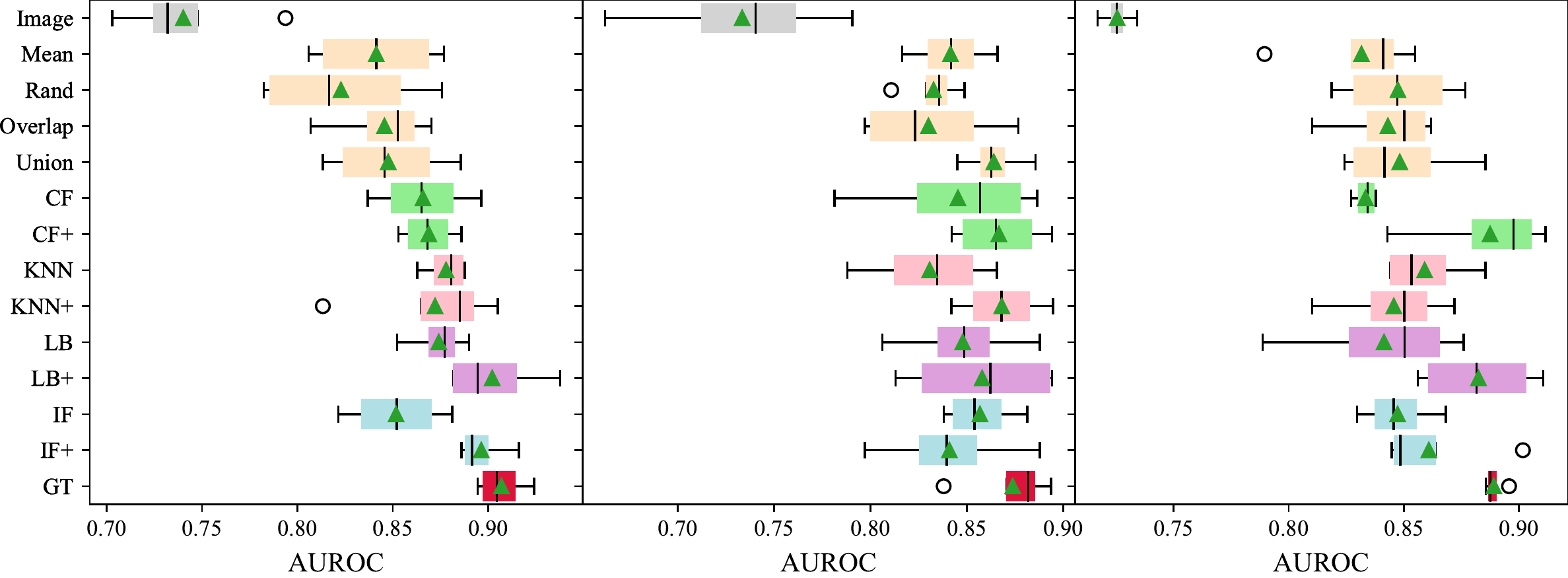}
			\put(11,-3.5){(a) Densenet 121~\cite{huang2017densely}}
			\put(44,-3.5){(b) Resnet 34~\cite{he2016deep}}	
			\put(75.5,-3.5){(c) Resnet 50}
		\end{overpic}
	\end{center}
	\caption{
		Quantitative comparisons of the diagnosis performance between binary masks derivatives and alpha mattes as input on three deep models, i.e., Densenet 121, Resnet 34, and Resnet 50. The results are shown in (a)-(c), respectively. 
		Mean, Rand, Overlap, and Union refer to the mean, randomly selected, overlap, and union of the manual labelled binary masks, respectively.
		The names with the suffix + stand for the corresponding optimized matting methods.
		The GT alpha matte obtains the best performance, and the optimized methods are generally better than their corresponding methods. Moreover, the matting methods generate better results than the binary mask derivatives due to the better ability to depict lesions.
		}
	\label{fig:diagnosis}
\end{figure}

\begin{figure}[!t]
	\begin{center}
		\begin{overpic}[width=\textwidth]{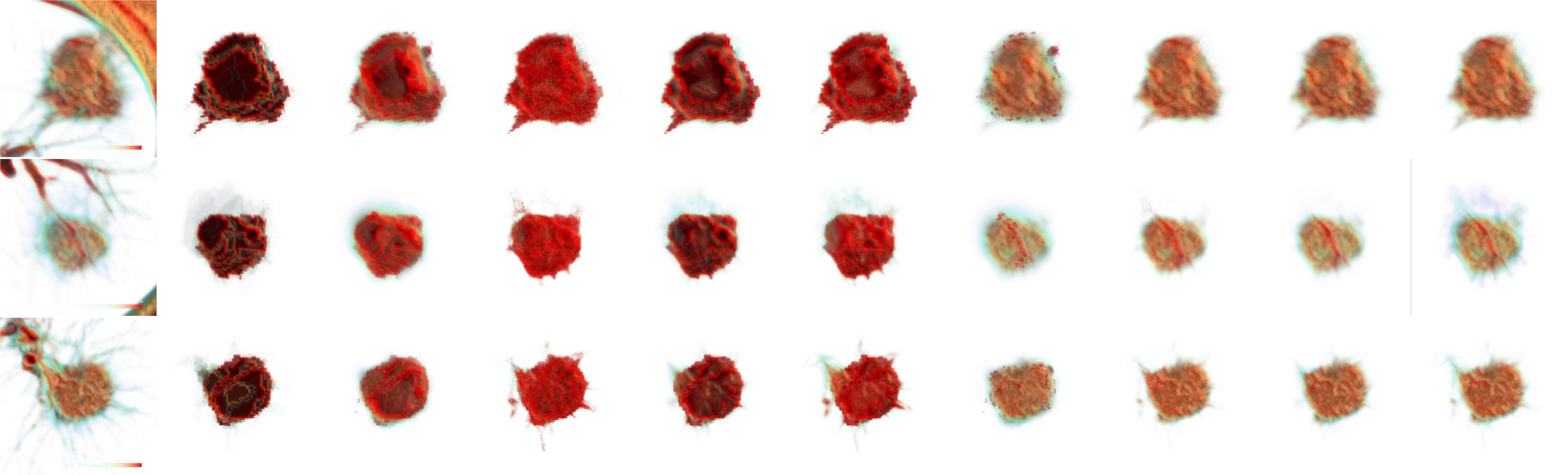}
			\put(0,-4.5){\scriptsize{(a) Image}}
			\put(10.5,-4.5){\scriptsize{(b) mMask}}	
			\put(22,-4.5){\scriptsize{(c) CF}}
			\put(30.5,-4.5){\scriptsize{(d) KNN}}
			\put(42,-4.5){\scriptsize{(e) LB}}
			\put(52.5,-4.5){\scriptsize{(f) IF}}
			\put(61.5,-4.5){\scriptsize{(g) CF+}}
			\put(70.5,-4.5){\scriptsize{(h) KNN+}}
			\put(82,-4.5){\scriptsize{(i) LB+}}
			\put(92.5,-4.5){\scriptsize{(j) IF+}}
		\end{overpic}
	\end{center}
	\caption{
		Different methods to describe the pulmonary nodules~\cite{armato2004lung}. 
		Original 3D images and the mean masks of the labelled binary masks are shown in (a) and (b). 
		We adapt four traditional state-of-the-art 2D matting methods to 3D, as shown in~(c)-(f). 
		The ability of these methods to represent the edges of the lesion is significantly improved over the binary masks, but the interior of the lesion is not numerically continuous with the edges. Therefore, we modify them by calibrating the foreground to the HU value in CT images, and the corresponding results are shown in (g)-(j). The continuity of the alpha mattes has been preserved.
		The videos in supplementary files provide more details.
		}
	\label{fig:traditional_alpha_mattes}
\end{figure}

The AUROCs of the models are shown in Figure~\ref{fig:diagnosis}. 
The GT alpha matte has the best AUROC, outperforming the traditional and the optimized methods, revealing that the created dataset effectively characterizes the nodular lesions. It also shows that our approach to constructing datasets semi-automatic is effective.
Also, in most cases, the alpha mattes perform better than the binary mask derivatives, which illustrates that the alpha mattes are more depictable than the binary masks. 
Moreover, the optimized methods also achieve better results than their counterparts because they can better describe the information inside of the lesions and could be a better substitute for the binary masks in lesion representation.
The original image ranks last, which may be related to its complicated input and the relatively small backbone of the classification network to capture effective features.

Figure~\ref{fig:traditional_alpha_mattes} shows the visualization of nodules and the alpha mattes generated by the traditional and the optimized methods. 
It reveals that the alpha mattes generated by the traditional methods give a better description of the lesion's edges than the binary masks, and preserve more diagnostic features (such as the ground-glass shadows). However, they are discontinuous at the edges and interior parts of lesions, which is inconsistent with the anatomical facts. 
In the optimized counterparts, the new alpha mattes have a more natural transition between the edges and interior parts of the lesions, while retaining the internal structural information. 

\subsection{Experiment on 3D Medical Matting Network}~\label{exp:3DMM}
\subsubsection{Implementation Details}
We deploy the proposed 3DMM on the annotated 3D matting dataset. 
The dataset is randomly divided into the training set, validation set, and test set, following the ratio of 7:1:2 according to the patient id to avoid possible interference between lesions in the same individual.
We augment the number of training samples by random cropping, flipping, and rotating.
Since the larger the input size, the more computational resources are required for training, we set the input size to $96\times 96 \times 96$ to achieve a good balance between performance and resource usage.
The widely used Adam optimizer is applied with weight decay set to $5\times 10^{-5}$~\cite{DBLP:journals/corr/KingmaB14}.
The cosine annealing learning rate policy is used after one epoch long steady increasing warm-up from $0$ to the base learning rate $3\times 10^{-4}$~\cite{loshchilov10sgdr,bochkovskiy2020yolov4,fang2020densely}. 
The loss weighting coefficients, i.e., $\eta$ and $\theta$, are set to $10$ through experimental searches.
We repeat the training four times, repartitioning the dataset each time to ensure the robustness of the results. All models are trained from scratch with batch-size $16$ and $300$ epochs to ensure the networks are fully converged. The Intel Xeon 6252 CPU and eight NVIDIA RTX 3090 GPUs are used in training.

	
\subsubsection{Results} We adapt the evaluation metrics widely used in image matting, namely, the SAD, MSE, Grad., and Conn., to the 3D scenes~\cite{rhemann2009perceptually}. 
The SAD, MSE, and Grad. are defined as:
\begin{equation}
	\mathit{SAD} (\hat{\boldsymbol{\alpha}},\boldsymbol{\alpha}) = \sum_{i \in \hat{\boldsymbol{\alpha}}} |\hat{\boldsymbol{\alpha}}_i-\boldsymbol{\alpha}_i|,
	\label{eq:SAD}
\end{equation}
\begin{equation}
	\mathit{MSE} (\hat{\boldsymbol{\alpha}},\boldsymbol{\alpha}) = \frac{1}{\sum \mathbbm{1}_{\hat{\boldsymbol{\alpha}}}} \sum_{i \in \hat{\boldsymbol{\alpha}}} \left(\hat{\boldsymbol{\alpha}}_i-\boldsymbol{\alpha}_i\right)^2,
	\label{eq:MSE}
\end{equation}
\begin{equation}
	\mathit{Grad} (\hat{\boldsymbol{\alpha}},\boldsymbol{\alpha})=
	\sum_{i \in \nabla_{\hat{\boldsymbol{\alpha}}}}
	\left ({\nabla_{\hat{\boldsymbol{\alpha}}}}_i-{\nabla_{\boldsymbol{\alpha}}}_i\right )^2,
	\label{eq:Grad}
\end{equation}
where $\hat{\boldsymbol{\alpha}}$ and $\boldsymbol{\alpha}$ stand for the predicted and GT alpha matte, respectively. $\nabla_{\hat{\boldsymbol{\alpha}}}$ represents the gradient of the predicted alpha matte. Conn. metrics the connectedness by means of connectivity in a set of binary threshold images of the evaluated alpha matte. We refer the readers to~\cite{rhemann2009perceptually} for more details.
SAD and MSE are computed to evaluate the difference between the predicted and the GT alpha matte, while Grad. and Conn. focus on assessing the continuity of the predictions. 

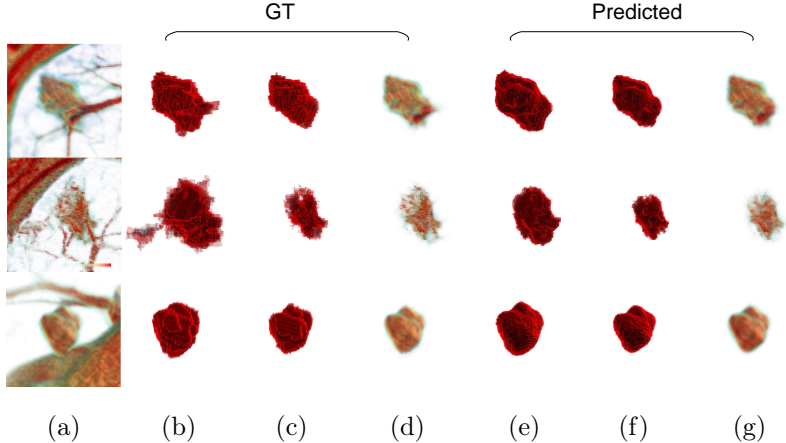
\begin{figure}[!t]
	\begin{center}
		\begin{overpic}[width=0.7\textwidth]{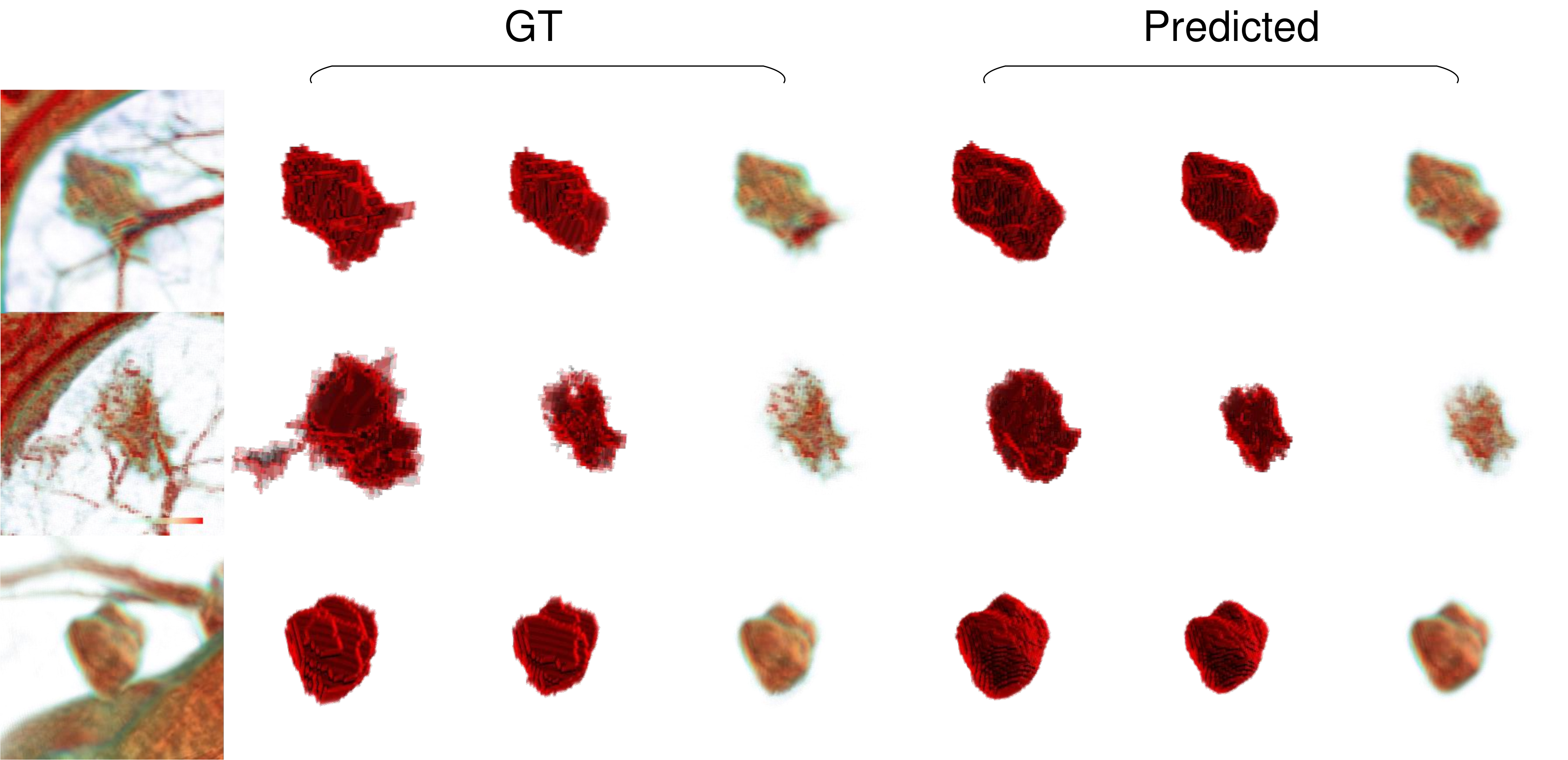}
			\put(5,-6){(a)}
			\put(19.25,-6){(b)}	
			\put(33.5,-6){(c)}
			\put(47.75,-6){(d)}
			\put(62.5,-6){(e)}
			\put(76.25,-6){(f)}
			\put(90.5,-6){(g)}
		\end{overpic}
	\end{center}
	\caption{Alpha mattes and masks predicted by 3D Medical Matting Network.
	The original image is shown in (a). The GT union masks, GT overlap masks, and GT alpha mattes are shown in (b)-(d), respectively. The predicted union masks, predicted overlap masks, and predicted alpha mattes are shown in (e)-(g), respectively.}
	\label{fig:dl_matting}
\end{figure}

\begin{table}[!t]
	\caption{Quantitative comparisons with 3D medical matting algorithms.
			The results are in the format of mean (± std). MSE is scaled by $1\times 10^{3}$, and SAD, Grad., and Conn. are scaled by $1\times 10^{-2}$. 
            The top three results are highlighted by \textbf{\color{green}{green}}, \textbf{\color{blue}{bule}}, and \textbf{\color{red}{red}}, respectively.}
	\centering
	\scriptsize
	\begin{threeparttable}
		\begin{tabular}{lc|cccc}
			\hline
			Model	&Fully Auto?	& SAD↓		& MSE↓		& Grad.↓	& Conn.↓	\\ \hline
			CF+\dag	&\usym{2717}	&152.62(±7.38)	&0.43(±0.03)	&14.96(±0.95)	&132.39(±7.10)	\\
			KNN+\dag	&\usym{2717}	&102.22(±6.68)	&0.25(±0.03)	&13.78(±1.18)	&76.01(±6.79)	\\
			LB+\dag	&\usym{2717}	&\textbf{\color{blue}{86.31(±15.52)}}	&\textbf{\color{green}{0.16(±0.05)}}	&\textbf{\color{green}{6.23(±1.45)}}	&\textbf{\color{green}{66.89(±14.80)}}	\\
			IF+\dag	&\usym{2717}	&\textbf{\color{green}{79.71(±8.68)}}	&\textbf{\color{blue}{0.18(±0.03)}}	&\textbf{\color{red}{7.88(±0.97)}}	&\textbf{\color{blue}{69.09(±8.17)}}	\\
			3DMM	&\usym{2713}	&\textbf{\color{red}{99.42(±6.42)}}	&\textbf{\color{red}{0.24(±0.02)}}	&\textbf{\color{blue}{6.37(±0.75)}}	&\textbf{\color{red}{69.25(±5.87)}}	\\ \hline
		\end{tabular}
		\begin{tablenotes}
			\item[\dag] \scriptsize{For fairness, the corresponding instances contributing to the GT alpha mattes are not counted in evaluation.}
		\end{tablenotes}
	\end{threeparttable}
	\label{tab:matting_results}
\end{table}

We compare the 3DMM with the optimized methods. 
Quantitative comparisons on the four matting metrics, as well as visualizations, are shown in Table~\ref{tab:matting_results} and Figure~\ref{fig:dl_matting}, respectively.
In Table~\ref{tab:matting_results}, it is shown that the proposed 3DMM ranks second in Grad. and Conn. metrics and third in SAD and MSE.
Despite the simplification of the network, 3DMM is comparable to the optimized methods in numerical results, which reveals that the DL-based approach is feasible for the 3D matting task. 
Moreover, the optimized methods take advantage of the prior information provided by the trimap, while 3DMM does not require additional human involvement. At the same time, it can easily implement parallel inference through a deep learning framework.

From Figure~\ref{fig:dl_matting}, the multi-task network achieves good results for the prediction of both masks and alpha matte and better expresses the details of the lesions. 
It is worth mentioning that because the DL-based methods learn potential rules from a set of data, some individual labelling errors have been corrected. The GT union mask of the second row in Figure~\ref{fig:dl_matting} illustrates an example. 

\subsection{Experiments on the efficient models}
In this part, the experiments to investigate the performances of efficient models are conducted. We evaluate the models with the matting metrics and computation resources.
Moreover, we verify the role of auxiliary masks by removing them from the matting branch as an ablation study.

\begin{figure}[!t]
	\begin{center}
		\includegraphics[width=0.8\textwidth]{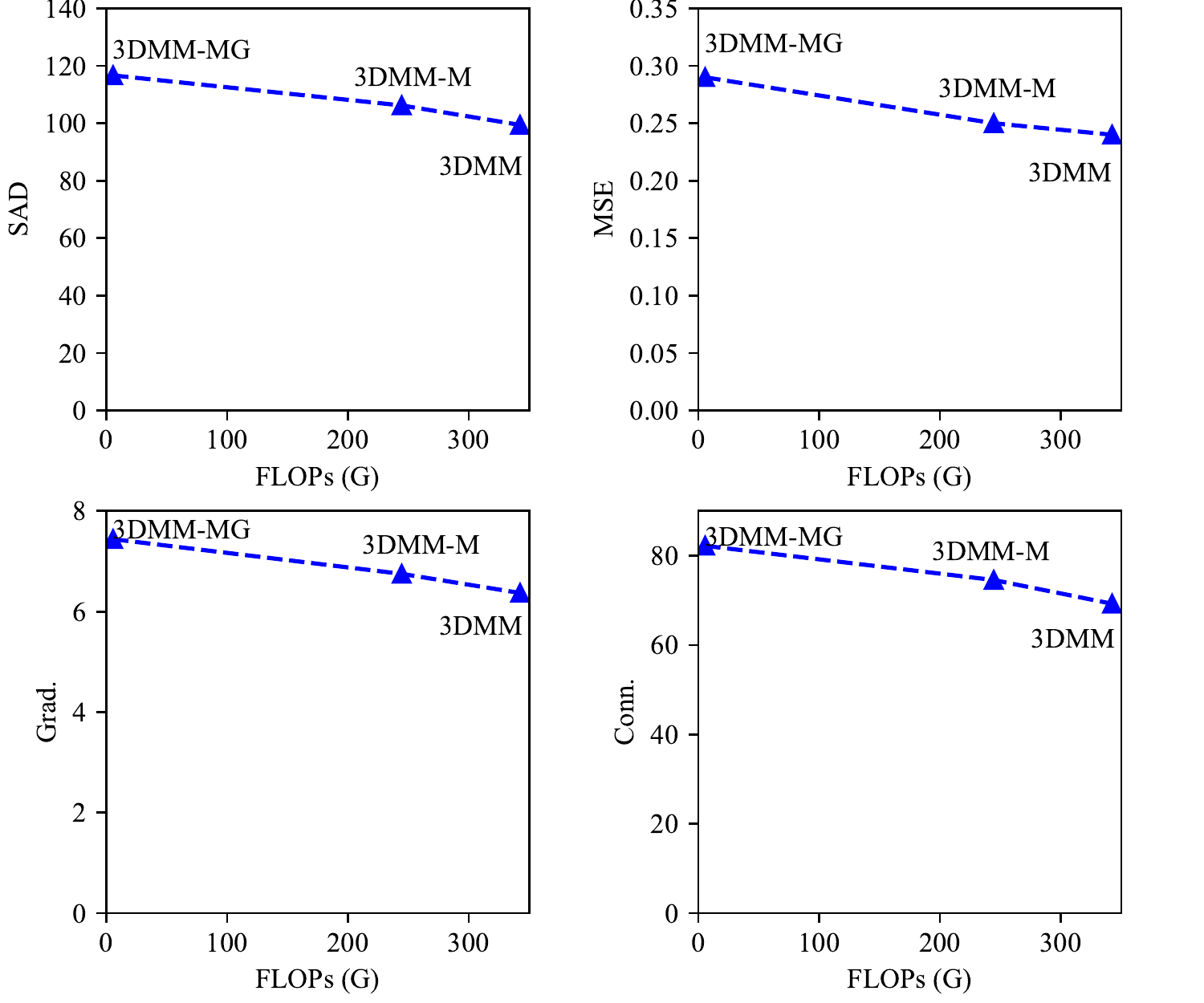}
	\end{center}
	\caption{
		Computation-performance trade-off of the efficient models.		
		The comparison of SAD, MSE, Grad., and Conn. metrics on the 3DMM, 3DMM-M, and 3DMM-MG models is shown from left to right.
	}
	\label{fig:flop_curve}
\end{figure}

In 3DMM, the filter numbers of the input, hidden layer, and output in each of the two consecutive convolutional layers unit or the ResBlock are equal, denoted as $n-n-n$. In the optimized model with ghost modules, the filter numbers of the Ghost ResBlock are $\frac{n}{4}-\frac{n}{2}-\frac{n}{4}$, and half of the filters of each ghost module are generated by the cheap transformations. The hyperparameters are introduced in Section~\ref{exp:3DMM}, except the learning rate for the models with ghost module is set to $3\times10^{-3}$ for better convergence. The matting performance and computation resources are compared in Table~\ref{tab:efficient_models} and Table~\ref{tab:computation_resources}, respectively. Figure~\ref{fig:flop_curve} shows the FLOPs-performance curves.

\begin{table}[!t]
	\caption{Quantitative comparisons with the DL-based 3D medical matting models.
	The results are in the format of mean (± std). MSE is scaled by $1\times 10^{3}$, and SAD, Grad., and Conn. are scaled by $1\times 10^{-2}$.}
	\centering
	\scriptsize
	\begin{threeparttable}
		\begin{tabular}{l|cccc}
			\hline
			Model	& SAD↓		& MSE↓		& Grad.↓	& Conn.↓	\\ \hline			
			3DMM	&\textbf{99.42(±6.42)}	&\textbf{0.24(±0.02)}	&\textbf{6.37(±0.75)}	&\textbf{69.25(±5.87)}	\\ 
			3DMM\dag	&106.33(±10.51)	&0.25(±0.04)	&6.75(±0.87)	&73.17(±9.24)	\\
			\hline
			3DMM-M	&\textbf{106.20(±8.67)}	&\textbf{0.25(±0.03)}	&\textbf{6.75(±0.59)}	&\textbf{74.56(±8.17)} \\
			3DMM-M\dag	&107.35(±8.60)	&0.26(±0.05)	&6.89(±0.91)	&76.34(±9.34)\\
			\hline
			3DMM-MG	&\textbf{116.66(±8.03)}	&\textbf{0.29(±0.04)}	&\textbf{7.44(±0.93)}	&\textbf{82.21(±8.41)} \\
			3DMM-MG\dag &123.80(±8.78)	&0.34(±0.07)	&8.87(±0.92)	&89.04(±9.63)	\\
			\hline
		\end{tabular}
		\begin{tablenotes}
   			\item[\dag] 3DMM without the auxiliary masks. 
      		\item[-M] 3DMM with the merged auxiliary branch. 
        	\item[-G] 3DMM with the Ghost modules.
		\end{tablenotes}
	\end{threeparttable}
	\label{tab:efficient_models}
\end{table}

\begin{table}[!t]
	\caption{Computation resources comparisons of the DL-based 3D medical matting models.
	The number of the network parameters and FLOPs are listed.}
	\centering
	\scriptsize
	\begin{threeparttable}
		\begin{tabular}{l|cccc}
			\hline
			Model	& Parameters		& FLOPs		\\ \hline			
			3DMM	& 2,940,327	& 342.46G\\ 
			3DMM-M	& 2,829,603	& 244.50G\\ 
			3DMM-MG	& 32,577	& 5.64G\\ 
			\hline
		\end{tabular}
	\end{threeparttable}
	\label{tab:computation_resources}
\end{table}

The model with the merged auxiliary branch takes 96.2\% parameters and 71.4\% FLOPs of the original 3DMM. By introducing the ghost module and reducing the filter numbers, the parameters and FLOPs are further reduced to 1.1\% and 1.6\% of the original 3DMM. The matting performance evaluated by SAD, MSE, Grad., and Conn. shows a slight decrease. However, compared to the significant computation reduction, the performance drop are acceptable. 
3D convolutional networks usually require an enormous number of computing resources. Therefore, efficient networks are important for 3D scenes. The results of the experiments illustrate the effectiveness of the proposed optimizations for efficient 3DMM.

The ablation study shown in Table~\ref{tab:efficient_models} illustrates that the proposed auxiliary mask mechanism can further improve the matting performance of the 3DMM and its high-efficiency counterparts.
In other words, using the binary segmentation results as a guide can effectively facilitate the matting task. 
This kind of experience allows other medical matting tasks to be quickly implementated. 
Moreover, the variance of the network with auxiliary masks is smaller than that of others networks, which indicates that the auxiliary mask mechanism is conducive to the models' robustness.

\section{Discussion and Limitation}\label{section:discussion}
Accurate lesion description is of great importance in the clinical setting. 
3D matting uses the soft mask to describe lesions with continuous values, which can provide more structural information than the binary mask and reflect the uncertainty of the lesions. 
Besides achieving better performance in the clinical diagnosis illustrated by the classification of pulmonary nodules in the paper, alpha matte may have promising applications in clinical treatment. For example, during radiotherapy, it is necessary to define the area of the lesion accurately. Alpha matte's soft labelling offers a richer possibility of dose design, providing different dosing schemes for lesion sections of different severity, which is conducive to personalized medicine.

Alpha matte can help us to better understand the relationship between the lesion and the original image. With this information, we can apply matting to the augmentation of medical images. For example, using alpha matte, lesions can be better fused into a new background, such as healthy tissues, to construct pseudo-positive cases closer to the real situation to alleviate the possible class imbalance during the training of deep learning networks.

Although we have explored 3D matting in medical scenes as comprehensively as possible, this research inevitably has some limitations.
For example, due to the lack of available datasets and annotations, we only conducted experiments on CT pulmonary nodules. Although our solution has methodological generality to other medical images (such as MRI), it has not been experimentally verified.
Also, we used a semi-automatic method in dataset construction, which made this research possible. However, it is still difficult to avoid a certain degree of subjectivity in dataset construction, and a more reliable and efficient method for dataset construction is the focus of our following research.

\section{Conclusions} \label{conclusion}
This work is the first comprehensive study of image matting in 3D medical scenes, which illustrates that alpha mattes are more expressive than binary masks in describing lesions qualitatively and quantitatively. 
We adapt four state-of-the-art matting methods in 2D to 3D scenes and customize these methods to better characterize the information inside the pulmonary nodules for CT images and show a better description at the edges than the binary masks. Based on these methods, the first 3D medical matting dataset is constructed efficiently. The validity of the alpha mattes labels is verified through the evaluation of clinicians and downstream experiments. The dataset will be released to the community. Meanwhile, a fully automatic matting method, 3DMM, is proposed to benchmark the study. It is the first study of the 3D DL-based matting method. The trimap-free property makes it more convenient to use, and the auxiliary branches provide binary segmentation masks at different confidence levels, which expand the application scope. The efficient models reduce the computational complexity without significant performance drop, which is suitable for equipment with limited computing resources.

\section*{Acknowledgments}
This work was supported by the National Natural Science Foundation of China [grant number 41876100, 42276187]; and the Fundamental Research Funds for the Central Universities [grant number 3072022FSC0401].

\bibliography{main.bbl}

\begin{thebibliography}{10}
\expandafter\ifx\csname url\endcsname\relax
  \def\url#1{\texttt{#1}}\fi
\expandafter\ifx\csname urlprefix\endcsname\relax\def\urlprefix{URL }\fi
\expandafter\ifx\csname href\endcsname\relax
  \def\href#1#2{#2} \def\path#1{#1}\fi

\bibitem{kohl2019hierarchical}
S.~A. Kohl, B.~Romera-Paredes, K.~H. Maier-Hein, D.~J. Rezende, S.~Eslami,
  P.~Kohli, A.~Zisserman, O.~Ronneberger, A hierarchical probabilistic {U-Net}
  for modeling multi-scale ambiguities, Medical Imaging meets NeurIPS Workshop
  (NeurIPS Workshop)\href {http://dx.doi.org/10.48550/arXiv.1905.13077}
  {\path{doi:10.48550/arXiv.1905.13077}}.

\bibitem{wang2021medical}
L.~Wang, L.~Ju, D.~Zhang, X.~Wang, W.~He, Y.~Huang, Z.~Yang, X.~Yao, X.~Zhao,
  X.~Ye, et~al., {Medical Matting}: a new perspective on medical segmentation
  with uncertainty, in: International Conference on Medical Image Computing and
  Computer-Assisted Intervention (MICCAI), Springer, 2021, pp. 573--583.
\newblock \href {http://dx.doi.org/10.1007/978-3-030-87199-4_54}
  {\path{doi:10.1007/978-3-030-87199-4_54}}.

\bibitem{li2021applications}
T.~Li, W.~Bo, C.~Hu, H.~Kang, H.~Liu, K.~Wang, H.~Fu, Applications of deep
  learning in fundus images: A review, Medical Image Analysis (MedIA) 69 (2021)
  101971.
\newblock \href {http://dx.doi.org/10.1016/j.media.2021.101971}
  {\path{doi:10.1016/j.media.2021.101971}}.

\bibitem{elbatel2022mammograms}
M.~Elbatel, Mammograms classification: A review, arXiv preprint
  arXiv:2203.03618\href {http://dx.doi.org/10.48550/arXiv.2203.03618}
  {\path{doi:10.48550/arXiv.2203.03618}}.

\bibitem{armato2004lung}
S.~G. Armato~III, G.~McLennan, M.~F. McNitt-Gray, C.~R. Meyer, D.~Yankelevitz,
  D.~R. Aberle, C.~I. Henschke, E.~A. Hoffman, E.~A. Kazerooni, H.~MacMahon,
  et~al., Lung image database consortium: developing a resource for the medical
  imaging research community, Radiology 232~(3) (2004) 739--748.
\newblock \href {http://dx.doi.org/10.1148/radiol.2323032035}
  {\path{doi:10.1148/radiol.2323032035}}.

\bibitem{codella2018skin}
N.~C. Codella, D.~Gutman, M.~E. Celebi, B.~Helba, M.~A. Marchetti, S.~W. Dusza,
  A.~Kalloo, K.~Liopyris, N.~Mishra, H.~Kittler, et~al., Skin lesion analysis
  toward melanoma detection: A challenge at the 2017 international symposium on
  biomedical imaging, in: IEEE 15th international symposium on biomedical
  imaging (ISBI), IEEE, 2018, pp. 168--172.
\newblock \href {http://dx.doi.org/10.48550/arXiv.1710.05006}
  {\path{doi:10.48550/arXiv.1710.05006}}.

\bibitem{kohl2018probabilistic}
S.~Kohl, B.~Romera-Paredes, C.~Meyer, J.~De~Fauw, J.~R. Ledsam, K.~Maier-Hein,
  S.~A. Eslami, D.~J. Rezende, O.~Ronneberger, A probabilistic {U-Net} for
  segmentation of ambiguous images, in: Advances in Neural Information
  Processing Systems (NeurIPS), 2018, pp. 6965--6975.
\newblock \href {http://dx.doi.org/10.48550/arXiv.1806.05034}
  {\path{doi:10.48550/arXiv.1806.05034}}.

\bibitem{baumgartner2019phiseg}
C.~F. Baumgartner, K.~C. Tezcan, K.~Chaitanya, A.~M. H{\"o}tker, U.~J.
  Muehlematter, K.~Schawkat, A.~S. Becker, O.~Donati, E.~Konukoglu, {PHiSeg}:
  Capturing uncertainty in medical image segmentation, in: International
  Conference on Medical Image Computing and Computer-Assisted Intervention
  (MICCAI), Springer, 2019, pp. 119--127.
\newblock \href {http://dx.doi.org/10.1007/978-3-030-32245-8_14}
  {\path{doi:10.1007/978-3-030-32245-8_14}}.

\bibitem{gantenbein2020revphiseg}
M.~Gantenbein, E.~Erdil, E.~Konukoglu, {RevPHiSeg}: A memory-efficient neural
  network for uncertainty quantification in medical image segmentation, in:
  Uncertainty for Safe Utilization of Machine Learning in Medical Imaging, and
  Graphs in Biomedical Image Analysis, Springer, 2020, pp. 13--22.
\newblock \href {http://dx.doi.org/10.1007/978-3-030-60365-6_2}
  {\path{doi:10.1007/978-3-030-60365-6_2}}.

\bibitem{dyer2020implications}
S.~C. Dyer, B.~J. Bartholmai, C.~W. Koo, Implications of the updated lung {CT}
  screening reporting and data system (lung-rads version 1.1) for lung cancer
  screening, Journal of Thoracic Disease 12~(11) (2020) 6966.
\newblock \href {http://dx.doi.org/10.21037/jtd-2019-cptn-02}
  {\path{doi:10.21037/jtd-2019-cptn-02}}.

\bibitem{zeng2012region}
Z.~Zeng, J.~Wang, T.~Shepherd, R.~Zwiggelaar, Region-based active surface
  modelling and alpha matting for unsupervised tumour segmentation in {PET},
  in: IEEE International Conference on Image Processing (ICIP), IEEE, 2012, pp.
  1997--2000.
\newblock \href {http://dx.doi.org/10.1109/icip.2012.6467280}
  {\path{doi:10.1109/icip.2012.6467280}}.

\bibitem{cheng2017awm}
J.~Cheng, M.~Zhao, M.~Lin, B.~Chiu, {AWM}: Adaptive weight matting for medical
  image segmentation, in: Medical Imaging 2017: Image Processing, Vol. 10133,
  International Society for Optics and Photonics, 2017, p. 101332P.
\newblock \href {http://dx.doi.org/10.1117/12.2254774}
  {\path{doi:10.1117/12.2254774}}.

\bibitem{zhao2020improving}
H.~Zhao, H.~Li, L.~Cheng, Improving retinal vessel segmentation with joint
  local loss by matting, Pattern Recognition (PR) 98 (2020) 107068.
\newblock \href {http://dx.doi.org/10.1016/j.patcog.2019.107068}
  {\path{doi:10.1016/j.patcog.2019.107068}}.

\bibitem{fan2018hierarchical}
Z.~Fan, J.~Lu, C.~Wei, H.~Huang, X.~Cai, X.~Chen, A hierarchical image matting
  model for blood vessel segmentation in fundus images, IEEE Transactions on
  Image Processing (TIP) 28~(5) (2018) 2367--2377.
\newblock \href {http://dx.doi.org/10.1109/TIP.2018.2885495}
  {\path{doi:10.1109/TIP.2018.2885495}}.

\bibitem{kim2021uacanet}
T.~Kim, H.~Lee, D.~Kim, {UACANet}: Uncertainty augmented context attention for
  polyp semgnetaion, Proceedings of the 29th ACM International Conference on
  Multimedia (2021) 2167--2175\href {http://dx.doi.org/10.1145/3474085.3475375}
  {\path{doi:10.1145/3474085.3475375}}.

\bibitem{khan2022deep}
S.~Khan, B.~Azam, Y.~Yao, W.~Chen, Deep collaborative network with alpha matte
  for precise knee tissue segmentation from mri, Computer Methods and Programs
  in Biomedicine 222 (2022) 106963.
\newblock \href {http://dx.doi.org/10.2139/ssrn.4040771}
  {\path{doi:10.2139/ssrn.4040771}}.

\bibitem{zhong20173d}
Z.~Zhong, Y.~Kim, J.~Buatti, X.~Wu, {3D} alpha matting based co-segmentation of
  tumors on {PET-CT} images, in: Molecular Imaging, Reconstruction and Analysis
  of Moving Body Organs, and Stroke Imaging and Treatment, Springer, 2017, pp.
  31--42.
\newblock \href {http://dx.doi.org/10.1007/978-3-319-67564-0_4}
  {\path{doi:10.1007/978-3-319-67564-0_4}}.

\bibitem{zhong2018improving}
Z.~Zhong, Y.~Kim, L.~Zhou, K.~Plichta, B.~Allen, J.~Buatti, X.~Wu, Improving
  tumor co-segmentation on {PET-CT} images with {3D} co-matting, in: IEEE 15th
  International Symposium on Biomedical Imaging (ISBI), IEEE, 2018, pp.
  224--227.
\newblock \href {http://dx.doi.org/10.1109/ISBI.2018.8363560}
  {\path{doi:10.1109/ISBI.2018.8363560}}.

\bibitem{liu2020three}
B.~Liu, X.~Zhang, L.~Yang, J.~Zhang, Three-dimensional organ extraction method
  for color volume image based on the closed-form solution strategy,
  Quantitative Imaging in Medicine and Surgery 10~(4) (2020) 862.
\newblock \href {http://dx.doi.org/10.21037/qims.2020.03.21}
  {\path{doi:10.21037/qims.2020.03.21}}.

\bibitem{levin2007closed}
A.~Levin, D.~Lischinski, Y.~Weiss, A closed-form solution to natural image
  matting, IEEE Transactions on Pattern Analysis and Machine Intelligence
  (TPAMI) 30~(2) (2007) 228--242.
\newblock \href {http://dx.doi.org/10.1109/TPAMI.2007.1177}
  {\path{doi:10.1109/TPAMI.2007.1177}}.

\bibitem{han2020ghostnet}
K.~Han, Y.~Wang, Q.~Tian, J.~Guo, C.~Xu, C.~Xu, Ghostnet: More features from
  cheap operations, in: Proceedings of the IEEE Conference on Computer Vision
  and Pattern Recognition (CVPR), 2020, pp. 1580--1589.
\newblock \href {http://dx.doi.org/10.1109/CVPR42600.2020.00165}
  {\path{doi:10.1109/CVPR42600.2020.00165}}.

\bibitem{kats2019soft}
E.~Kats, J.~Goldberger, H.~Greenspan, Soft labeling by distilling anatomical
  knowledge for improved ms lesion segmentation, in: 2019 IEEE 16th
  International Symposium on Biomedical Imaging (ISBI 2019), IEEE, 2019, pp.
  1563--1566.
\newblock \href {http://dx.doi.org/10.1109/isbi.2019.8759518}
  {\path{doi:10.1109/isbi.2019.8759518}}.

\bibitem{dai2022soft}
P.~Dai, L.~Dong, R.~Zhang, H.~Zhu, J.~Wu, K.~Yuan, Soft-cp: A credible and
  effective data augmentation for semantic segmentation of medical lesions,
  arXiv preprint arXiv:2203.10507\href
  {http://dx.doi.org/10.48550/arXiv.2203.10507}
  {\path{doi:10.48550/arXiv.2203.10507}}.

\bibitem{aksoy2017designing}
Y.~Aksoy, T.~Ozan~Aydin, M.~Pollefeys, Designing effective inter-pixel
  information flow for natural image matting, in: Proceedings of the IEEE
  Conference on Computer Vision and Pattern Recognition (CVPR), 2017, pp.
  29--37.
\newblock \href {http://dx.doi.org/10.1109/CVPR.2017.32}
  {\path{doi:10.1109/CVPR.2017.32}}.

\bibitem{cai2019disentangled}
S.~Cai, X.~Zhang, H.~Fan, H.~Huang, J.~Liu, J.~Liu, J.~Liu, J.~Wang, J.~Sun,
  Disentangled image matting, in: Proceedings of the IEEE International
  Conference on Computer Vision (ICCV), 2019, pp. 8819--8828.
\newblock \href {http://dx.doi.org/10.1109/ICCV.2019.00891}
  {\path{doi:10.1109/ICCV.2019.00891}}.

\bibitem{chen2013knn}
Q.~Chen, D.~Li, C.-K. Tang, {KNN} {Matting}, IEEE Transactions on Pattern
  Analysis and Machine Intelligence (TPAMI) 35~(9) (2013) 2175--2188.
\newblock \href {http://dx.doi.org/10.1109/TPAMI.2013.18}
  {\path{doi:10.1109/TPAMI.2013.18}}.

\bibitem{chuang2001bayesian}
Y.-Y. Chuang, B.~Curless, D.~H. Salesin, R.~Szeliski, A bayesian approach to
  digital matting, in: Proceedings of the IEEE Conference on Computer Vision
  and Pattern Recognition (CVPR), Vol.~2, IEEE, 2001, pp. II--264--II--271.
\newblock \href {http://dx.doi.org/10.1109/cvpr.2001.990970}
  {\path{doi:10.1109/cvpr.2001.990970}}.

\bibitem{forte2020fbamatting}
M.~Forte, F.~Piti{\'e}, {F, B, Alpha Matting}, arXiv preprint
  arXiv:2003.07711\href {http://dx.doi.org/10.48550/arXiv.2003.07711}
  {\path{doi:10.48550/arXiv.2003.07711}}.

\bibitem{lutz2018alphagan}
S.~Lutz, K.~Amplianitis, A.~Smolic, Alphagan: Generative adversarial networks
  for natural image matting, British Machine Vision Conference (BMVC)\href
  {http://dx.doi.org/10.48550/arXiv.1807.10088}
  {\path{doi:10.48550/arXiv.1807.10088}}.

\bibitem{wang2008image}
J.~Wang, M.~F. Cohen, et~al., Image and video matting: a survey, Foundations
  and Trends{\textregistered} in Computer Graphics and Vision 3~(2) (2008)
  97--175.
\newblock \href {http://dx.doi.org/10.1561/9781601981356}
  {\path{doi:10.1561/9781601981356}}.

\bibitem{xu2017deep}
N.~Xu, B.~Price, S.~Cohen, T.~Huang, Deep image matting, in: Proceedings of the
  IEEE Conference on Computer Vision and Pattern Recognition (CVPR), 2017, pp.
  2970--2979.
\newblock \href {http://dx.doi.org/10.1109/cvpr.2017.41}
  {\path{doi:10.1109/cvpr.2017.41}}.

\bibitem{yao2017comprehensive}
G.~Yao, Z.~Zhao, S.~Liu, A comprehensive survey on sampling-based image
  matting, in: Computer Graphics Forum, Vol.~36, Wiley Online Library, 2017,
  pp. 613--628.
\newblock \href {http://dx.doi.org/10.1111/cgf.13156}
  {\path{doi:10.1111/cgf.13156}}.

\bibitem{chen2022pp}
G.~Chen, Y.~Liu, J.~Wang, J.~Peng, Y.~Hao, L.~Chu, S.~Tang, Z.~Wu, Z.~Chen,
  Z.~Yu, et~al., {PP-Matting}: High-accuracy natural image matting, arXiv
  preprint arXiv:2204.09433\href {http://dx.doi.org/10.48550/arXiv.2204.09433}
  {\path{doi:10.48550/arXiv.2204.09433}}.

\bibitem{2016Deep}
X.~Shen, X.~Tao, H.~Gao, C.~Zhou, J.~Jia, Deep automatic portrait matting, in:
  European Conference on Computer Vision (ECCV), Springer, 2016, pp. 92--107.
\newblock \href {http://dx.doi.org/10.1007/978-3-319-46448-0_6}
  {\path{doi:10.1007/978-3-319-46448-0_6}}.

\bibitem{zheng2009learning}
Y.~Zheng, C.~Kambhamettu, Learning based digital matting, in: Proceedings of
  the IEEE International Conference on Computer Vision (ICCV), IEEE, 2009, pp.
  889--896.
\newblock \href {http://dx.doi.org/10.1109/iccv.2009.5459326}
  {\path{doi:10.1109/iccv.2009.5459326}}.

\bibitem{weisstein2002kronecker}
E.~W. Weisstein,
  \href{https://mathworld.wolfram.com/KroneckerDelta.html}{Kronecker delta},
  https://mathworld. wolfram. com/.
\newline\urlprefix\url{https://mathworld.wolfram.com/KroneckerDelta.html}

\bibitem{denotter2021hounsfield}
T.~D. DenOtter, J.~Schubert,
  \href{https://europepmc.org/article/NBK/nbk547721}{Hounsfield unit} (2021).
\newline\urlprefix\url{https://europepmc.org/article/NBK/nbk547721}

\bibitem{ronneberger2015u}
O.~Ronneberger, P.~Fischer, T.~Brox, U-net: Convolutional networks for
  biomedical image segmentation, in: International Conference on Medical image
  computing and computer-assisted intervention (MICCAI), Springer, 2015, pp.
  234--241.
\newblock \href {http://dx.doi.org/10.1007/978-3-319-24574-4_28}
  {\path{doi:10.1007/978-3-319-24574-4_28}}.

\bibitem{he2016deep}
K.~He, X.~Zhang, S.~Ren, J.~Sun, Deep residual learning for image recognition,
  in: Proceedings of the IEEE Conference on Computer Vision and Pattern
  Recognition (CVPR), 2016, pp. 770--778.
\newblock \href {http://dx.doi.org/10.1109/cvpr.2016.90}
  {\path{doi:10.1109/cvpr.2016.90}}.

\bibitem{kopuklu2019resource}
O.~K{\"o}p{\"u}kl{\"u}, N.~Kose, A.~Gunduz, G.~Rigoll, Resource efficient 3d
  convolutional neural networks, in: IEEE International Conference on Computer
  Vision Workshop (ICCVW), IEEE, 2019, pp. 1910--1919.
\newblock \href {http://dx.doi.org/10.1109/ICCVW.2019.00240}
  {\path{doi:10.1109/ICCVW.2019.00240}}.

\bibitem{krizhevsky2012imagenet}
A.~Krizhevsky, I.~Sutskever, G.~E. Hinton, Imagenet classification with deep
  convolutional neural networks, Advances in neural information processing
  systems (NeurIPS) 25.
\newblock \href {http://dx.doi.org/10.1145/3065386}
  {\path{doi:10.1145/3065386}}.

\bibitem{huang2017densely}
G.~Huang, Z.~Liu, L.~Van Der~Maaten, K.~Q. Weinberger, Densely connected
  convolutional networks, in: Proceedings of the IEEE Conference on Computer
  Vision and Pattern Recognition (CVPR), 2017, pp. 4700--4708.
\newblock \href {http://dx.doi.org/10.1109/cvpr.2017.243}
  {\path{doi:10.1109/cvpr.2017.243}}.

\bibitem{DBLP:journals/corr/KingmaB14}
D.~P. Kingma, J.~Ba, \href{http://arxiv.org/abs/1412.6980}{Adam: {A} method for
  stochastic optimization}, in: Y.~Bengio, Y.~LeCun (Eds.), 3rd International
  Conference on Learning Representations, {ICLR} 2015, San Diego, CA, USA, May
  7-9, 2015, Conference Track Proceedings, 2015.
\newline\urlprefix\url{http://arxiv.org/abs/1412.6980}

\bibitem{loshchilov10sgdr}
I.~Loshchilov, F.~Hutter, {SGDR}: {S}tochastic gradient descent with warm
  restarts, The International Conference on Learning Representations
  (ICLR)\href {http://dx.doi.org/10.48550/arXiv.1608.03983}
  {\path{doi:10.48550/arXiv.1608.03983}}.

\bibitem{bochkovskiy2020yolov4}
A.~Bochkovskiy, C.-Y. Wang, H.-Y.~M. Liao, Yolov4: Optimal speed and accuracy
  of object detection, arXiv preprint arXiv:2004.10934\href
  {http://dx.doi.org/10.48550/arXiv.2004.10934}
  {\path{doi:10.48550/arXiv.2004.10934}}.

\bibitem{fang2020densely}
J.~Fang, Y.~Sun, Q.~Zhang, Y.~Li, W.~Liu, X.~Wang, Densely connected search
  space for more flexible neural architecture search, in: Proceedings of the
  IEEE Conference on Computer Vision and Pattern Recognition (CVPR), 2020, pp.
  10628--10637.
\newblock \href {http://dx.doi.org/10.48550/arXiv.1906.09607}
  {\path{doi:10.48550/arXiv.1906.09607}}.

\bibitem{rhemann2009perceptually}
C.~Rhemann, C.~Rother, J.~Wang, M.~Gelautz, P.~Kohli, P.~Rott, A perceptually
  motivated online benchmark for image matting, in: Proceedings of the IEEE
  Conference on Computer Vision and Pattern Recognition (CVPR), IEEE, 2009, pp.
  1826--1833.
\newblock \href {http://dx.doi.org/10.1109/cvpr.2009.5206503}
  {\path{doi:10.1109/cvpr.2009.5206503}}.

\end{thebibliography}

\end{document}